\documentclass[twocolumn,aps,prd,showpacs, nofootinbib]{revtex4-2}
\pdfoutput=1
\usepackage{amsmath,amsfonts}
\usepackage{epsfig}
\usepackage{caption}
\usepackage{subcaption}
\usepackage{ifpdf}
\usepackage{hyperref}
\usepackage{amsmath}
\usepackage{graphicx}
\usepackage{color}
\usepackage{natbib}
\usepackage{multirow}
\usepackage{kotex}
\usepackage{times}

\newcommand{\deff}{D_{\rm eff}}
\newcommand{\msun}{M_\odot}
\newcommand{\be}{\begin{equation}}
\newcommand{\ee}{\end{equation}}
\newcommand{\bea}{\begin{eqnarray}}
\newcommand{\eea}{\end{eqnarray}}

\newcommand{\etal}{{\it et al.}}

\begin{document}

\title{Systematic bias due to eccentricity in parameter estimation for merging binary neutron stars}

\author{Hee-Suk Cho}
\email{chohs1439@pusan.ac.kr}
\affiliation{Department of Physics, Pusan National University, Busan, 46241, Korea}
\affiliation{Extreme Physics Institute, Pusan National University Busan, 46241, Korea}

\date{\today}

\begin{abstract}

We study the impact of eccentricity on gravitational-wave parameter estimation for binary neutron star systems.
For signals with small eccentricity injected into noise with the advanced LIGO power spectral density, 
we perform Bayesian parameter estimation using the circular waveform model
and show how the recovered parameters can be biased from their true values, 
focusing on the intrinsic parameters the chirp mass ($M_c$), the symmetric mass ratio ($\eta$), and the tidal deformability ($\tilde{\lambda}$).
By comparing the results between the Bayesian and the analytic Fisher-Cutler-Vallisneri (FCV) methods, 
we obtain the valid criteria for the FCV approach.
Employing the FCV method and using the realistic population of binary neutron star sources distributed in the $m_1$--$m_2$--$e_0$ space, 
where $e_0$ indicates the eccentricity at 10Hz,
we calculate the measurement errors ($\sigma_{\theta}$) and the systematic biases ($\Delta \theta/\sigma_{\theta}$)
and obtain their generalized distributions in the range of $0 \leq e_0 \leq 0.025$.
We find that for all of the three parameters, the biases increase with increasing $e_0$,
and this increase is faster for larger $e_0$.
The bias is mainly dependent on the value of $e_0$
and weakly dependent on the component masses, and thus the distribution shows a narrow band in the $e_0$--$\Delta \theta/\sigma_{\theta}$ plane.
We present various posterior examples to illustrate our new findings, such as the bimodality of posteriors.
In particular, we give a specific injection-recovery example to 
demonstrate the importance of including eccentricity in parameter estimation to avoid incorrect predictions of the neutron star equation of state.

\end{abstract}


\maketitle

\section{Introduction}
Recently, the ground-based gravitational-wave (GW) observatories, advanced LIGO (aLIGO) \cite{ALIGO} and advanced Virgo \cite{AVirgo},
have finished the third observing run.
The network of the two aLIGO detectors and the advanced Virgo detector has detected $90$
GW candidates \cite{GWTC-1,GWTC-2,GWTC-2.1,GWTC-3},
all GW sources have been verified to be merging compact binaries.
Most of the GW signals were emitted from the binary black holes (BBHs),
and a small number of signals were emitted from two binary neutron stars (BNSs)
and a few neutron star-black hole (NSBH) binaries.
In particular, the neutron star (NS) tidal parameter could be measured from 
the first detected BNS signal GW170817 due to its sufficiently high signal-to-noise ratio \cite{GW170817,GW170817PE}.

So far, various waveform models have been developed for use in GW data analysis.
The search template bank is constructed in the three-dimensional parameter space
consisting of the component masses and the aligned-spin \cite{PhysRevD.99.024048,PhysRevD.103.084047}.
In this pipeline, an inspiral-only post-Newtonian (PN) waveform model is used for BNS templates \cite{PhysRevD.80.084043}
and a full inspiral-merger-ringdown (IMR) waveform model is used for BBH and NSBH templates \cite{PhysRevD.95.044028}.
In parameter estimation, more diverse models are used to infer source properties
as accurately as possible, taking into account the effect of waveform systematics.
Various IMR models defined in the time domain or frequency domain have been used for BBH signals, 
some of which include the precession effect and/or the higher multipoles (for example, see Table I of \cite{PhysRevD.102.043015}).
For BNS signals, an inspiral-only PN model and several IMR models have been used,
and these models included NS tidal effects \cite{GW170817PE,GW190425}.

On the other hand, the orbital eccentricity has not been taken into account 
in the current LIGO-Virgo GW searches since the binary system is likely to be circularized when 
they reach the detector frequency band ($\sim 10$Hz) as the systems can radiate
away their eccentricity through GWs \cite{PhysRev.131.435,PhysRev.136.B1224}.
However, it has been known that 
dynamically formed binaries in dense stellar environments may still have
non-negligible eccentricities ($e>0.1$) at 10Hz \cite{PhysRevLett.120.151101,PhysRevD.97.103014}.
Therefore, the presence of a dynamical formation channel
is expected to be confirmed by successful measurements of the nonsmall eccentricity.
In this context, searches for eccentric BBHs \cite{Abbott_2019} or BNSs \cite{Nitz_2020} 
have also been performed using LIGO-Virgo data, but no new significant candidates have been identified.
Several groups have attempted to estimate the eccentricity of the BBH sources detected by LIGO-Virgo
using the inspiral-only \cite{10.1093/mnras/staa1176}, the aligned-spin \cite{10.1093/mnras/stz2996,Romero_Shaw_2021,OShea:2021ugg}, 
and the numerical simulation \cite{Gayathri:2022tj} waveform models, respectively.
The authors of \cite{10.1093/mnrasl/slaa084,10.1093/mnras/staa2120} also tried to measure the eccentricity of the BNS sources
separately, and both obtained very small eccentricities. 
There have also been studies on the search efficiency for eccentric BBHs \cite{PhysRevD.93.043007,PhysRevD.97.024031}  
or eccentric BNSs \cite{PhysRevD.87.127501} using quasicircular templates.

Studies on the impact of eccentricity on GW parameter estimation
have also been done in several works.
The authors of \cite{PhysRevD.92.044034,Gond_n_2018,Gond_n_2019} found that
eccentricity can improve the measurement accuracy of mass parameters.
Recently, some injection studies have been performed in \cite{PhysRevD.98.083028,PhysRevD.105.023003}
to investigate the measurability of eccentricity by ground-based detectors. 
In particular, a systematic parameter bias that can be produced by using a quasicircular waveform model for eccentric BBH signals
has been investigated in the recent works \cite{OShea:2021ugg} and \cite{PhysRevD.105.023003}.
They showed that the eccentric signals can be recovered by the quasicircular binaries with higher chirp masses ($M_c$) and higher symmetric mass ratios ($\eta$).
The two works \cite{PhysRevLett.112.101101} and \cite{PhysRevD.92.044034} have also conducted research on the bias due to eccentricity in BNS parameter estimation.
The former used the analytic method introduced by Cutler and Vallisneri \cite{PhysRevD.76.104018} to calculate the systematic bias, while
the latter calculated the bias by searching the $M_c$ -- $\eta$ overlap surface.
Both considered the equal-mass BNS system with $m_i=1.4\msun$ but showed different results on the bias for $\eta$.
In the former work, as the signal's eccentricity increases, the bias increases, exceeding the statistical error at small eccentricities of $\sim 0.002$, 
while in the latter work, the bias can be ignored compared to the statistical error even at large eccentricities of $\sim 0.4$.
As we will see in our results, such a difference is due to the fact that 
since the overlap surface is restricted by the physical boundary (i.e., $0 < \eta \leq 0.25$),
the recovered value of $\eta$ cannot exceed 0.25, resulting in negligible bias.
However, these boundary effects are not considered in the analytic method,
and thus the recovered $\eta$ can increase beyond 0.25.

Unlike the case for the BBH system, the waveform of the BNS system has additional parameters related to the NS tidal deformability.
Although the tidal parameters have a very small effect compared to other parameters such as masses or spins, 
the tidal deformability is a very important source property because 
this parameter is related to the internal structure of NSs.
Precision measurements of tidal deformability can directly constrain the NS equation of state (EOS) \cite{PhysRevD.81.123016,PhysRevD.85.123007,PhysRevLett.111.071101,PhysRevD.92.023012,Harry_2018}.
The tidal deformability was directly measured from GW170817 for the first time,
and as a result, the soft EOS models were preferred over stiff EOS models \cite{GW170817PE,GW170817EOS}.
Meanwhile, studies on the impact of eccentricity on the measurement of tidal deformability
have not yet been conducted except for the simple analytic results presented by Favata \cite{PhysRevLett.112.101101}.  

In this work, we study the impact of eccentricity on parameter estimation for BNS signals.
For injection signals with various small eccentricities $e_0$, where $e_0$ indicates the eccentricity
of the binary at the GW frequency of 10 Hz,
we perform Bayesian parameter estimation using circular waveforms
and investigate the biases for the intrinsic parameters the chirp mass, the symmetric mass ratio, and the tidal deformability.
We also use the Fisher matrix-based analytic method to calculate the bias
and evaluate the validity of the analytic method by comparing the results with the Bayesian parameter estimation results.
By applying the analytic method to the $10^4$ Monte Carlo samples randomly generated in the $m_1$--$m_2$--$e_0$ space, 
we obtain the general distributions of bias for the parameters.
In particular, we provide a variety of posterior examples for ease of understanding.

\section{Method}

\subsection{Waveform model: TaylorF2}
So far, many waveform models have been proposed
for use in GW data analysis.
In contrast to the BBH systems that require full IMR waveforms,
it is a large advantage for the BNS systems that
 inspiral-only waveform models can be used efficiently in the ground-based detectors
 because the portion of the postmerger is negligible
 as it goes out of the detector's frequency band.
 This can be directly confirmed from the data analysis results for GW170817 \cite{GW170817PE} and GW190425 \cite{GW190425},
 which showed consistent results between the IMR and the inspiral-only waveform models.
Therefore, we adopt the TaylorF2 waveform model in this work.
Since TaylorF2 is derived from the time-domain PN waveform model using a stationary phase approximation,
this model is defined in the frequency domain.
TaylorF2 is well formulated by the PN expansion with the corresponding higher-order corrections
and has a simpler form than other models used in current GW data analysis \cite{PhysRevD.95.044028, PhysRevD.98.084028,PhysRevD.101.124040, PhysRevD.102.044055,PhysRevD.89.084006,PhysRevD.95.024010, PhysRevLett.120.161102, PhysRevD.100.024059, 
PhysRevD.101.024056, PhysRevD.99.064045, PhysRevLett.113.151101,PhysRevLett.116.181101,P_rrer_2014,PhysRevD.93.044007,PhysRevLett.113.151101}

The wave function of TaylorF2 is given by
\be \label{eq.TaylorF2}
h(f)={M_{\rm c}^{5/6} \over \pi ^{2/3} \deff } \sqrt{{5 \over 24}} f^{-7/6} e^{i \Psi(f)},
\ee
where $\deff$ indicates the effective distance defined by a combination of the five extrinsic parameters;
those are the true distance of the source ($d_L$), orbital inclination with respect to the line of sight ($\theta_{JN}$), sky position ($RA, DEC$),
and polarization angle ($\psi$) \cite{PhysRevD.85.122006}.  
The wave phase of TaylorF2 is give by
\bea \label{eq.TaylorF2phase}
\Psi(f)&=&2\pi ft_c -2 \phi_c -{\pi\over 4} \\     \nonumber
 &+& {3 \over 128 \eta v^5} [\phi^{\rm pp,circ}(f)+ \phi^{\rm tidal}(f)+\phi^{\rm ecc}(f)],
\eea
where $v\equiv[\pi f (m_1+m_2)]^{1/3}$ is the PN expansion parameter and    
$t_c$ and $\phi_c$ are the coalescence time and phase, respectively.
The term $\phi^{\rm pp,circ}$ represents the point-particle, circular PN expansion,
we here use the standard 3.5PN equation given in \cite{PhysRevD.79.104023} 
but do not consider the spin effect.
The term $\phi^{\rm tidal}$ represents the tidal corrections.
Although the PN tidal corrections first appear in the 5PN order, its impact is comparable to the 3.5PN order point-particle correction \cite{PhysRevD.77.021502}.
The PN tidal corrections are given as a function of the two dimensionless tidal deformability $\lambda_1$ and $\lambda_2$.
We consider the tidal corrections up to 6PN order, then the tidal term can be given by the two useful parameters $\tilde{\lambda}$ and $\delta \tilde{\lambda}$ as \cite{PhysRevLett.112.101101,PhysRevD.91.043002}
\be \label{eq.tidal correction}
 \phi^{\rm Tidal} = - \bigg[ \frac{39 \tilde{\lambda}}{2}v^{10} + \bigg( \frac{3115\tilde{\lambda}}{64} - \frac{6595 \sqrt{1-4\eta} \delta \tilde{\lambda}}{364} \bigg)v^{12} \bigg].
\ee
In this case, the 5PN order term is only a function of $\tilde{\lambda}$,
so the contribution of $\delta \tilde{\lambda}$ can be greatly reduced in the overall tidal effect.
As we will see, $\delta \tilde{\lambda}$ has a negligible influence on the parameter estimation results.
Finally, the term $\phi^{\rm ecc}$ represents the eccentricity corrections,
we here consider the PN eccentricity corrections derived up to 3PN order given in Eq. (6.26) of \cite{PhysRevD.93.124061},
which can be expressed as
\bea
\phi^{\rm ecc}&=&-\frac{2355}{1462} e_0^2 \left(\frac{v_0}{v}\right)^{19/3} \bigg[1+\bigg(\frac{18766963}{2927736}\eta \\   \nonumber
&+&\frac{299076223}{81976608}\bigg) v^2 +\left(\frac{2833}{1008}-\frac{197}{36} \eta \right) v_0^2  \\   \nonumber
&+& O(v^4)+...+O(v^6)\bigg],
\eea
where $e_0$ is the eccentricity at the reference frequency $f_0$ (where the choice of $f_0$ is arbitrary).
For the ground-based detectors, the above eccentric PN waveform model is valid up to $e_0 \lesssim 0.1$ for comparable mass systems \cite{PhysRevD.93.124061}.

Therefore, in order to produce the TaylorF2 waveforms for eccentric and nonspinning BNS systems, we need the five extrinsic parameters ($d_L, \theta_{JN},  \psi, RA, DEC$), 
the five intrinsic parameters ($M_c, \eta, \tilde{\Lambda}, \delta \tilde{\Lambda}, e_0$),
and the two arbitrary constants ($t_c, \phi_c$), hence a total of 12 parameters are required.

\subsection{Bayesian parameter estimation}
The main purpose of the GW search pipeline is to identify
GW signals buried in the detector's data stream \cite{PhysRevD.85.122006}.
A more detailed analysis is conducted in the parameter estimation pipeline
to determine the source properties by exploring the entire parameter space \cite{PhysRevD.91.042003}.
Using Bayesian inference statistics, the result of parameter estimation can be 
given as posterior probability density functions (PDFs) for the parameters considered. 
The Bayesian parameter estimation process is based on iterative computations of the overlap 
between the model waveforms ($h$) and the detector data ($x$) containing the signal waveform ($s$).
The overlap between $x$ and $h$ is given by
\be \label{eq.overlap}
\langle x | h \rangle = 4 {\rm Re} \int_{f_{\rm min}}^{f_{\rm max}}  \frac{\tilde{x}(f)\tilde{h}^*(f)}{S_n(f)} df,
\ee
where the tilde denotes the Fourier transform of the time-domain
waveform, $S_n(f)$ is the detector's noise power spectral density (PSD),
$f_{\rm min}$ and $f_{\rm max}$ are the minimum and the maximum cutoff frequencies, respectively. 
For a given prior distribution $p(\theta)$, where $\theta$ is the set of parameters considered in the analysis,
the posterior probability that the GW signal is characterized by the parameters $\theta$
is expressed as
\be \label{eq.posterior}
p(\theta|x) \propto p(\theta)L(x|\theta), 
\ee
where $L(x|\theta)$ indicates the likelihood function.
Using the overlap defined in Eq. (\ref{eq.overlap}), 
the likelihood can be given by \cite{PhysRevD.46.5236,PhysRevD.49.2658}
\be \label{eq.likelihood}
L(x|\theta) \propto {\rm exp}\bigg[-\frac{1}2 \langle x-h(\theta)|x-h(\theta)\rangle\bigg].
\ee
In the limit of high SNRs where the noise effect can be ignored,
the above equation can be expressed as
\be \label{eq.likelihood-2}
L(\theta)\propto {\rm exp} \bigg[ -\frac{1}2\{\langle s|s \rangle+\langle h(\theta)|h(\theta)\rangle-2\langle s|h(\theta)\rangle\} \bigg].
\ee

For a given signal $s$, the SNR is given by \cite{PhysRevD.85.122006}
\be \label{eq.snr}
\rho=\sqrt{\langle s|s \rangle}.
\ee
If we use an accurate waveform model, the likelihood distribution has a maximum value
when the parameter values of the model waveform are equal to those of the signal waveform
(i.e., $h(\theta_0) \simeq s$, hence $\rho^2=\langle s|s \rangle \simeq \langle h(\theta_0)|h(\theta_0) \rangle$).
In this case, around the maximum position, the likelihood distribution in Eq. (\ref{eq.likelihood-2}) can be given as \cite{PhysRevD.87.024004}
\be \label{eq.likelihood-3}
L(\theta) \propto {\rm exp} [-\rho^2\{1- \langle \hat{s}|\hat{h}(\theta)\rangle\}],
\ee
where $\hat{h} \equiv h/ \rho$ denotes the normalized waveform.
Therefore, in Bayesian parameter estimation, 
the shape of the {\bf log}--likelihood is determined by the overlap between the signal and the model waveforms,
and its scale of interest depends on the SNR \cite{PhysRevD.87.024004}.

Since the prior information is just applied at the beginning of the process,
the construction of the parameter estimation algorithm is a matter of finding the global maximum of the likelihood distribution over the entire parameter space
as quickly and accurately as possible.
In this work, we perform Bayesian parameter estimation for BNS injection signals using the {\bf Bilby} library \cite{Ashton_2019}, 
which is one of the parameter estimation packages,
with the {\bf Dynesty} \cite{dynesty:Speagle_2020} nested sampling algorithm.
In particular, to speed up the parameter estimation runs, we apply the multibanded likelihood technique described in \cite{PhysRevD.104.044062}.
Furthermore, we select the phase marginalization option for efficiency, which enables analytical marginalization over $\phi_c$.

\subsection{Fisher matrix approach: Parameter measurement error and systematic bias}

In the high SNR limit, since the likelihood distribution obeys the Gaussian approximately,
the marginalized 1--d posterior PDF for the parameter can also be expressed as a Gaussian function \cite{PhysRevD.46.5236,PhysRevD.49.2658}.
The standard deviation of the PDF corresponds to the measurement error of the corresponding parameter.
On the other hand, for a given signal with the true parameter values $\theta_0$,
the measurement errors can be easily obtained by using the Fisher matrix defined by
\be\label{eq.FM}
\Gamma_{ij} = \bigg \langle {\partial h(\theta) \over \partial \theta_i} \bigg| {\partial h(\theta) \over \partial \theta_j} \bigg \rangle\bigg|_{\theta=\theta_0}.
\ee
The inverse of Fisher matrix gives the covariance matrix ($\Sigma_{ij} = (\Gamma^{-1})_{ij} $) of the parameter errors, 
and the measurement error ($\sigma_i$) of each parameter and the correlation coefficient ($C_{ij}$) between two parameters are given by
\be \label{eq.sigma}
\sigma_i=\sqrt{\Sigma_{ii}},  \ \ \ C_{ij}={\Sigma_{ij} \over \sqrt{\Sigma_{ii} \Sigma_{jj}}}.
\ee
Despite some well-known limitations (for more details, refer to \cite{PhysRevD.77.042001}),
the Fisher matrix method has been widely used in the
parameter estimation literature 
since it was first used by \cite{PhysRevD.49.2658,PhysRevD.52.848}.

Another formula for the Fisher matrix can be given by the likelihood function as \cite{PhysRevD.49.1723,PhysRevD.77.042001}
\be\label{eq.FM-2}
\Gamma_{ij} = - \frac{\partial^2 {\rm ln} L(\theta)}{\partial \theta_i \partial \theta_j} \bigg|_{\theta=\theta_0}.
\ee
Applying the likelihood function in Eq. (\ref{eq.likelihood-3}) to the above equation gives
\be \label{eq.FM-3}
\Gamma_{ij}=-\rho^2 {\partial^2   \langle \hat{s}|\hat{h}(\theta)\rangle \over \partial \theta_i \partial \theta_j}\bigg|_{\theta=\theta_0},
\ee
and this represents the curvature of the overlap surface at the maximum position.
Thus, one can infer that specific isomatch contours in the overlap surface
correspond to the confidence regions of the posterior PDF for a given SNR \cite{PhysRevD.87.024035,Cho_2014,Cho_2015}.

One of the limitations of the Fisher matrix method is the applicability of prior information.
Unlike Bayesian parameter estimation, only Gaussian prior functions can be imposed analytically on the Fisher matrix.
Assuming that the variance of the Gaussian prior function for the parameter $\theta_i$ is $P^2_{\theta_i}$,
the covariance matrix containing the prior information can be given by adding the component  $\Gamma_{ii}^0=P_{\theta_i}^{-2}$ as \cite{PhysRevD.49.2658,PhysRevD.52.848}
\be  \label{eq.covariant_matrix_with_prior}
\Sigma^P_{ij} = (\Gamma_{ij} + \Gamma_{ii}^0)^{-1},
\ee
where $ \Gamma_{ij}$ is the original Fisher matrix with no prior information.
The prior is independent of the signal strength, while the original Fisher matrix strongly depends on the SNR as seen in Eq. (\ref{eq.FM-3}).
Therefore, the lower the SNR, the stronger the prior effect on the parameter measurement.

Based on the Fisher matrix formalism, Cutler and Vallisneri \cite{PhysRevD.76.104018} 
developed an analytic method (henceforth denoted as ``FCV" following the notation in \cite{PhysRevD.105.023003})
to calculate the systematic bias.
In this work, the bias can be induced by ignoring the eccentricity in the model waveforms in the analysis of the eccentric signals. 
Following Eq. (29) of \cite{PhysRevD.76.104018}, the systematic bias can be given by \cite{PhysRevLett.112.101101}
\be
\Delta \theta_i=\theta^{\rm rec}_i-\theta_{i0}=\Sigma_{ij}(\partial_j h_{\rm AP}|h_{\rm T} - h_{\rm AP}),
\ee
where $\theta^{\rm rec}$ represents the recoverd parameter value, $h_{\rm T}$ indicates the true signal waveform, 
and $h_{\rm AP}$ indicates the approximate waveform in which the eccentricity corrections are ignored.
Here, $\Sigma_{ij}$ should be calculated from $h_{\rm AP}$.
By representing the wave amplitude of Eq. (\ref{eq.TaylorF2}) as $A$, the above equation can be rewritten as 
\be \label{eq.bias}
\Delta \theta_i=4 A^2 \Sigma_{ij} \int_{f_{\rm min}}^{f_{\rm max}} df  \frac{f^{-7/3}}{S_n(f)}(\Psi_{\rm T}-\Psi_{\rm AP})\partial_j \Psi_{\rm AP} ,
\ee
where $\Psi_{\rm T}$  indicates the true wave phase given in Eq. (\ref{eq.TaylorF2phase}) and $\Psi_{\rm AP}$ indicates the approximate phase. 
Since $A^2 \propto \rho^2$ and $\Sigma_{ij} \propto \rho^{-2}$, the bias is independent of the SNR.
To include the prior effect in the bias computation, the prior-incorporated covariance matrix ($\Sigma^P_{ij}$) can be used in this equation.

\section{Result}

\subsection{Setup}
The eccentric TaylorF2 waveform model is implemented in LAL (LIGO Algorithm Library) \cite{lalsuite} named as ``TaylorF2Ecc".
Using this model, we generate eccentric and nonspinning BNS signals
and inject those into the LIGO-Hanford detector with zero noise.
We use the aLIGO sensitivity curve labeled as ``aligo$\_$O4high" \footnote{https://dcc.ligo.org/LIGO-T2000012/public},
and the minimum frequency cutoff is set to $f_{\rm min}=20$Hz. 
The maximum frequency cutoff is set to $f_{\rm max}=f_{\rm ISCO}$,
where $f_{\rm ISCO}$ is the innermost-stable-circular-orbit frequency defined by $f_{\rm ISCO}=1/[6^{3/2}\pi (m_1+m_2)]$.
We choose the fiducial BNS masses as $(m_1, m_2)=(2\msun, 1\msun)$,
and then the tidal parameters can be given as ($\lambda_1,\lambda_2)=(14.7,1744)$.
Here, we assume the APR4 model \cite{PhysRevD.79.124032}, which is one of the soft EOS models 
preferred by the parameter estimation results for GW170817 \cite{GW170817PE}.

When running {\bf Bilby},
the algorithm must explore the entire 11--d parameter space.
However, we can reduce the exploration space to 6--d space
by holding the extrinsic parameters equal to their true values.
Note that, including the extrinsic parameters in the analysis has a little impact on the results of the intrinsic parameters (see, Appendix \ref{ap.full-param}).
We assume the flat priors in the ranges of
$[0.5\msun, \ 2.5\msun]$ for the component masses ($m_1,m_2$) and [0,  5000] for the component tidal parameters ($\lambda_1,\lambda_2$).
The priors of $t_c$ and $\phi_c$ are given as $[t_{c0}-1s,t_{c0}+1s]$ and [$0,2\pi$], respectively.
Since our parameter estimation is performed with circular waveforms, the eccentricity is not included in the prior setting.
When displaying the posteriors in {\bf Bilby}, one can choose the parameters of interest,
and we select ($\tilde{\lambda},\delta \tilde{\lambda})$ rather than $(\lambda_1, \lambda_2)$ in our results.

When computing the Fisher matrix,
we use the circular waveform model that can be obtained by ignoring the eccentric term in Eq. (\ref{eq.TaylorF2phase}).
Thus, the input parameters of the Fisher matrix are given as ($M_c,\eta,\tilde{\lambda},\delta \tilde{\lambda}, t_c,\phi_c$).
Although the choice of $t_c$ and $\phi_c$ does not affect the results of the intrinsic parameters,
they should be considered with the intrinsic parameters, as they usually have strong correlations with the intrinsic parameters.
We adopt a single-detector configuration and only focus on the intrinsic parameters in this work.
The extrinsic parameters only scale the signal strength, 
and most of the information about the intrinsic parameters is contained in the wave phase. 
Therefore, we do not consider the extrinsic parameters by choosing a fixed effective distance ($\deff$)
as in the case of Bayesian parameter estimation.
If $\deff$ is used as one of the input parameters in the Fisher matrix, 
its correlations with the intrinsic parameters are relatively very small \cite{Cho_2022}.

\subsection{Measurement error: Comparison between Bayesian and Fisher matrix methods}

\begin{figure*}[t]
\begin{center}
\includegraphics[width=2\columnwidth]{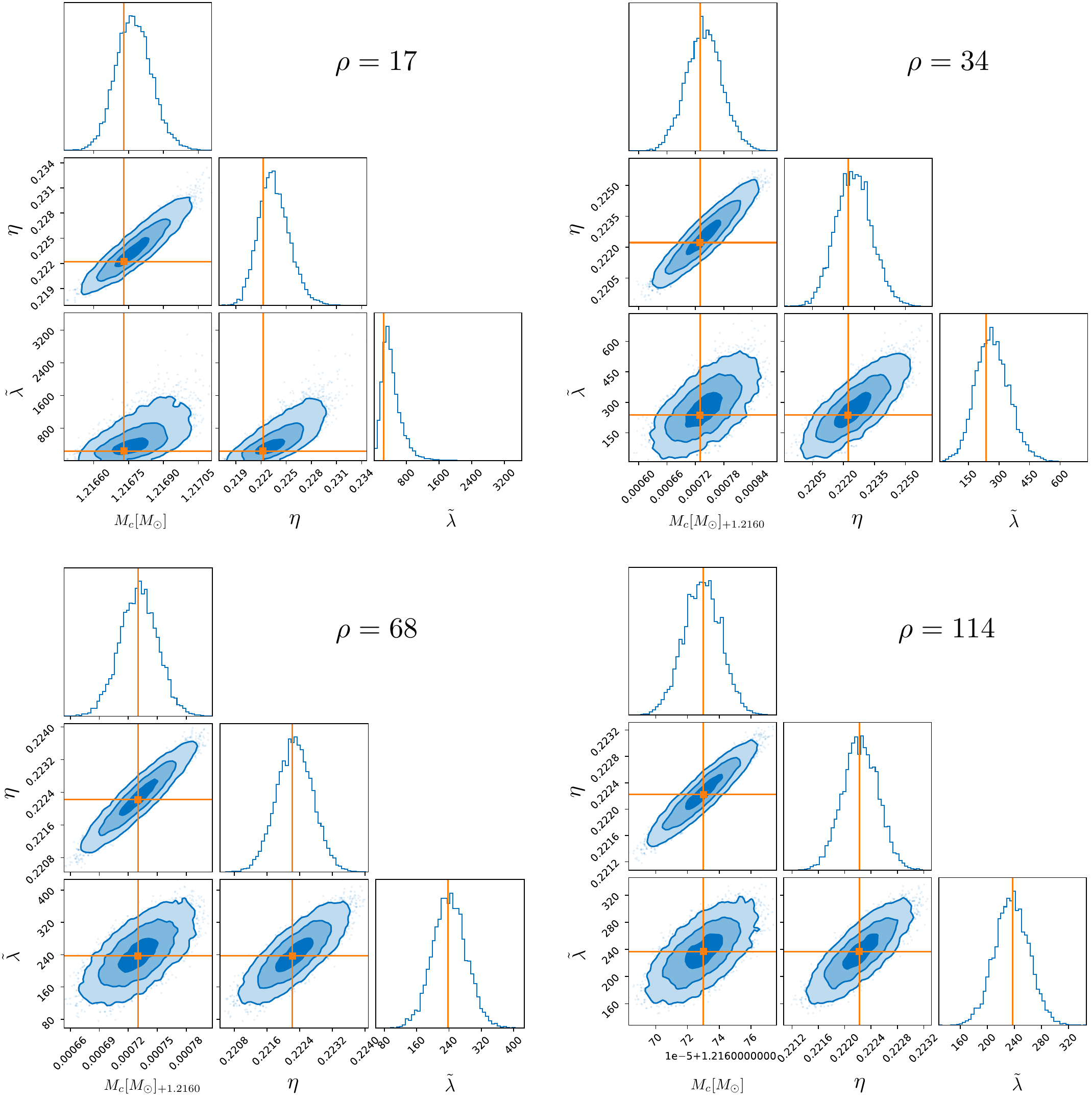}
\caption{\label{fig.bilby-errors}Marginalized 1--d posterior PDF and 2--d confidence regions with the SNRs $\rho=17, 34, 68,$ and $114$.
The contours indicate 39, 86, and $99 \%$ confidence regions. Injection values are 
$(m_1,m_2, [M_c,\eta], \lambda_1,\lambda_2, [\tilde{\lambda}, \delta \tilde{\lambda}])=(2\msun, 1\msun, [1.21673\msun, 0.22222],14.7,1744, [237,101])$ marked in orange.}
\end{center}
\end{figure*}

\begin{figure}[t]
\begin{center}
\includegraphics[width=\columnwidth]{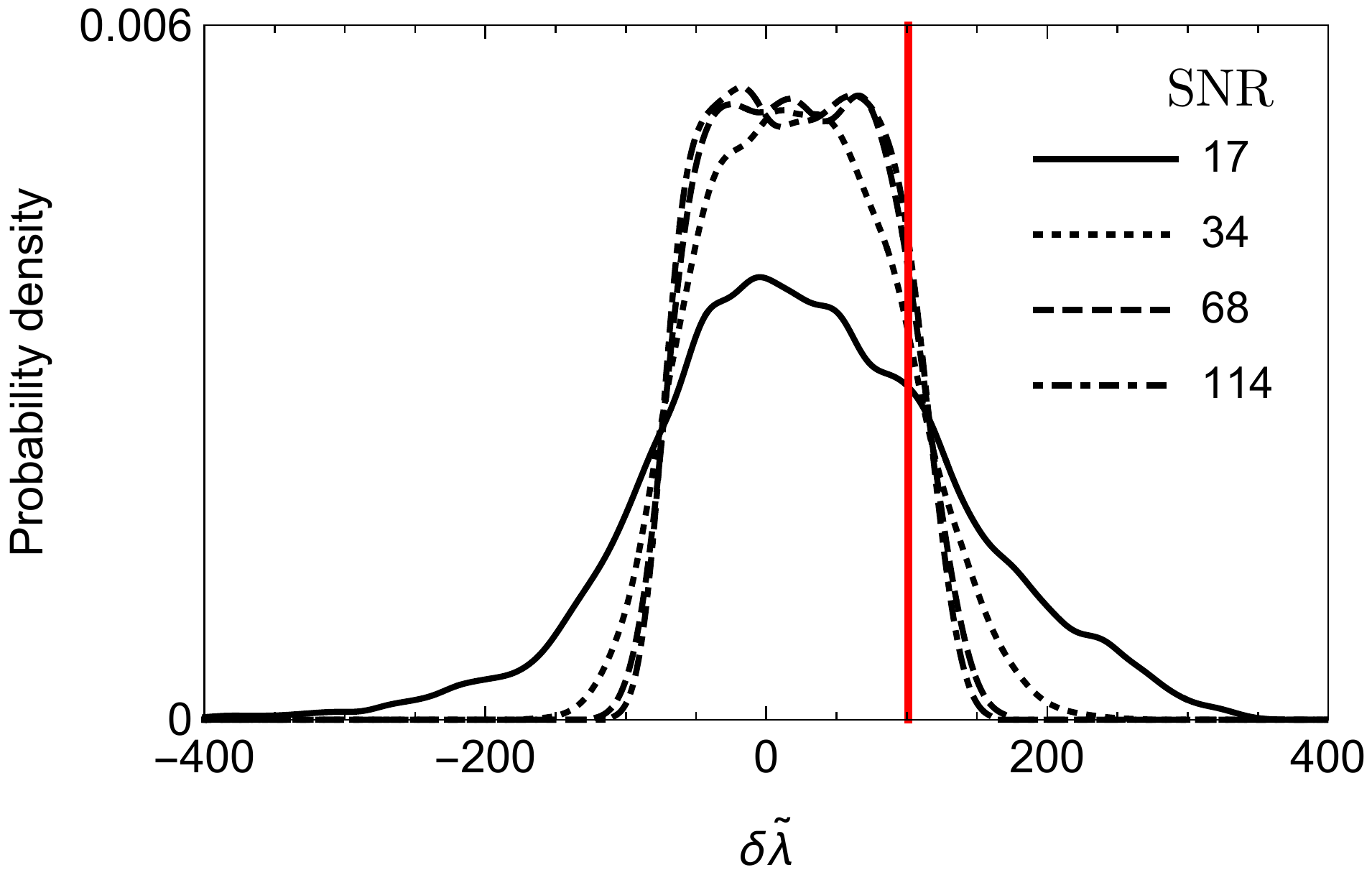}
\caption{\label{fig.error-delta} Marginalized PDFs for $\delta \tilde{\lambda}$. The true value is marked in red.}
\end{center}
\end{figure}

First, we perform parameter estimation using circular waveforms for a circular signal with our fiducial parameter values assuming various SNRs.
In this case, in principle, the posterior should be unbiased 
because we use the same waveform model as the signal model and assume zero noise in the injections.
Figure \ref{fig.bilby-errors} shows the results with the SNRs of (17, 34, 68, 114), where the true values are marked in orange.
The SNRs can be simply obtained by choosing $\deff=(200, 100, 50,30)$ Mpc.
For each SNR, the marginalized 1--d posterior PDF and the 2--d confidence regions are shown for the parameters $M_c, \eta,$ and $\tilde{\lambda}$.
It can be seen that the injection signals can be recovered well overall, showing decreasing measurement errors with increasing SNR.
The PDFs are slightly biased from the true values for low SNRs, but rarely biased for high SNRs.
For $\rho=17$, the contours are cut by the physical boundary for $\tilde{\lambda}$ and asymmetric around the true value,
while the contours are clearly symmetric for high SNRs. 
The marginalized PDF for $\delta \tilde{\lambda}$ is given in Fig. \ref{fig.error-delta} separately.
Unlike the three parameters given in Fig. \ref{fig.bilby-errors}, 
$\delta \tilde{\lambda}$ is poorly recovered for all SNRs.
In particular, the PDFs for $\rho=68$  and 114 are very similar, and they show no Gaussian distribution.

\begin{figure}[t]
\begin{center}
\includegraphics[width=\columnwidth]{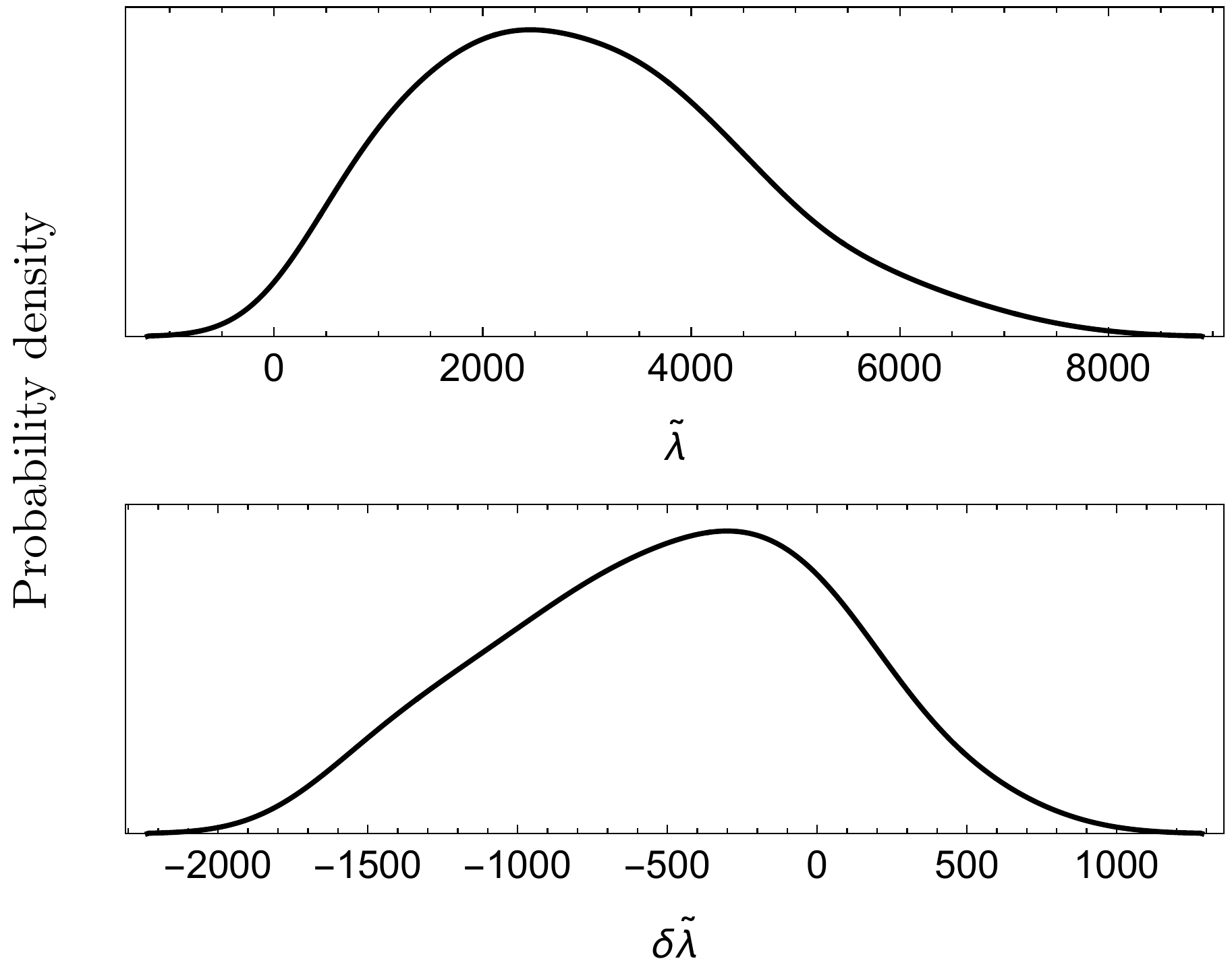}
\caption{\label{fig.lambda-delta-distribution}Prior distributions for $\tilde{\lambda}$ (upper) and $\delta \tilde{\lambda}$ (lower) used in Bayesian parameter estimation. 
These can be obtained by converting the flat priors for $(m_1,m_2,\lambda_1,\lambda_2)$ in the ranges of 
$[0.5\msun, \ 2.5\msun]$ for the component masses and [0,  5000] for the component tidal parameters.}
\end{center}
\end{figure}

Next, by applying the circular waveform model to the Fisher matrix method,
we calculate the measurement errors assuming no prior information.
The Fisher matrix is given as a 6--d matrix with the parameter components $(M_c, \eta, \tilde{\lambda}, \delta \tilde{\lambda},t_c, \phi_c)$,
and we use the same parameter values as in the previous Bayesian parameter estimation.
By  comparing the results between the Fisher matrix and the Bayesian posterior,
we find that the measurement errors from the Fisher matrix are much larger than
those from the Bayesian parameter estimation, especially for the tidal parameters.

\begin{figure}[t]
\begin{center}
\includegraphics[width=\columnwidth]{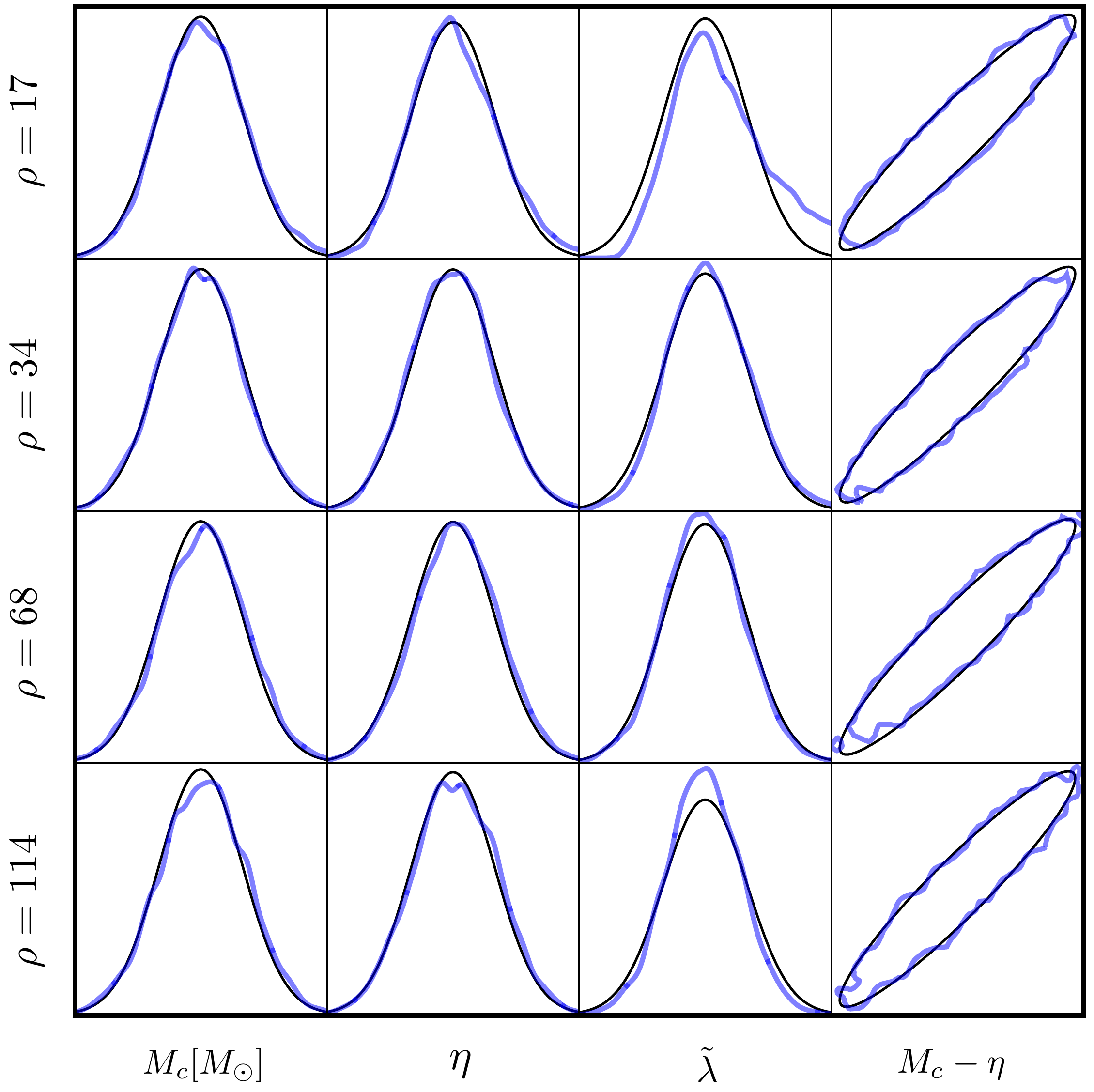}
\caption{\label{fig.bilbyvsFM}Comparison between the Fisher matrix (black) and Bayesian parameter estimation (blue). 
The 2--d contour indicates $68\%$ confidence region given in the $M_c$--$\eta$ plane. 
Note that, the small biases are not shown here since the Bayesian posteriors are shifted slightly to match the Gaussian functions.}
\end{center}
\end{figure}

The discrepancy between the Fisher matrix and the parameter estimation
is mainly due to the prior effect.
Figure \ref{fig.lambda-delta-distribution} shows the prior distributions for $\tilde{\lambda}$ and $\delta \tilde{\lambda}$ used in the Bayesian parameter estimation,
which can be obtained by converting our prior distributions for $(m_1,m_2,\lambda_1,\lambda_2)$.
It can be seen that the posterior PDFs for $\delta \tilde{\lambda}$ in Fig. \ref{fig.error-delta} 
are significantly reduced compared to the prior in Fig. \ref{fig.lambda-delta-distribution}, as expected.
However, we found that the measurement error of $\delta \tilde{\lambda}$ obtained from the Fisher matrix 
was much larger than the width of the Bayesian prior for $\delta \tilde{\lambda}$ in Fig. \ref{fig.lambda-delta-distribution}.
Motivated by that, we impose the Gaussian prior for the parameter $\delta \tilde{\lambda}$, whose variance is similar to the 
Bayesian prior size ($P_{\delta \tilde{\lambda}}\sim 500$), to the Fisher matrix.
Using the prior-incorporated Fisher matrix,
we calculate the measurement errors ($\sigma$) of $M_c, \eta,$ and $\tilde{\lambda}$ and their correlation coefficients.
Then, we compare the 1--d Gaussian functions determined by $\sigma$ assuming $\rho=17,34,68$, and $114$ with the marginalized Bayesian PDFs shown in Fig. \ref{fig.bilby-errors}.
In Fig. \ref{fig.bilbyvsFM}, the black and the blue lines correspond to the Fisher matrix and the Bayesian results, respectively, where
the contour indicates $68 \%$ confidence region given in the $M_c$--$\eta$ plane. 
This result clearly shows that the analytic Gaussian functions agree remarkably well with the Bayesian PDFs for all SNRs.
The confidence regions are also nearly identical between the two methods, indicating that the correlations are also very similar.  
Note that, the Bayesian posteriors are shifted slightly to match the Gaussian functions, so the small biases are not shown here.
In Table \ref{tab.FM-error}, we list the measurement error and correlation coefficient for the parameters $M_c, \eta,\tilde{\lambda},$ and $\delta\tilde{\lambda}$ assuming $\rho=34$.
The result with no prior information is given in the top row, and the result with the prior $P_{\delta \tilde{\lambda}}\sim 500$ is given in the middle row.

\begin{table}[b]
\centering
\begin{tabular}{c|cccc|ccc}
\hline\hline
$P_{\delta \tilde{\lambda}}$&$ \sigma_{M_c}/M_c$ & $\sigma_{\eta}$ & $\sigma_{\tilde{\lambda}}$  & $\sigma_{\delta \tilde{\lambda}}$ & $C_{M_c \delta \tilde{\lambda}}$ &$C_{\eta \delta \tilde{\lambda}}$&$C_{\tilde{\lambda} \delta \tilde{\lambda}}$  \\
\hline
- & 4.521E-5  & 1.739E-3  & 990.1  &  19780 & 0.7271  &0.8607&0.9965\\
500 & 3.105E-5 & 0.886E-3  & 87.0  &  499.8 & 0.0268  &0.0426&0.2865\\
-  & 3.104E-5 & 0.886E-3  & 83.4  & -& -&-&-\\
 \hline\hline
\end{tabular}
\caption{Measurement errors ($\sigma_i$) and correlation coefficients ($C_{ij}$) calculated by the Fisher matrix method assuming $\rho=34$. 
The top row is obtained from the 6--d Fisher matrix with no prior information. 
The middle row is obtained from the 6--d Fisher matrix incorporating the Gaussian prior on the parameter 
$\delta \tilde{\lambda}$ with the variance ($P_{\delta \tilde{\lambda}}\sim 500$) that is similar to the 
Bayesian prior size shown in Fig. \ref{fig.lambda-delta-distribution}. The bottom row is obtained from the 5--d Fisher matrix where $\delta \tilde{\lambda}$ is not considered.}
\label{tab.FM-error}
\end{table}

\begin{figure}[t]
    \centering
        \includegraphics[width=\columnwidth]{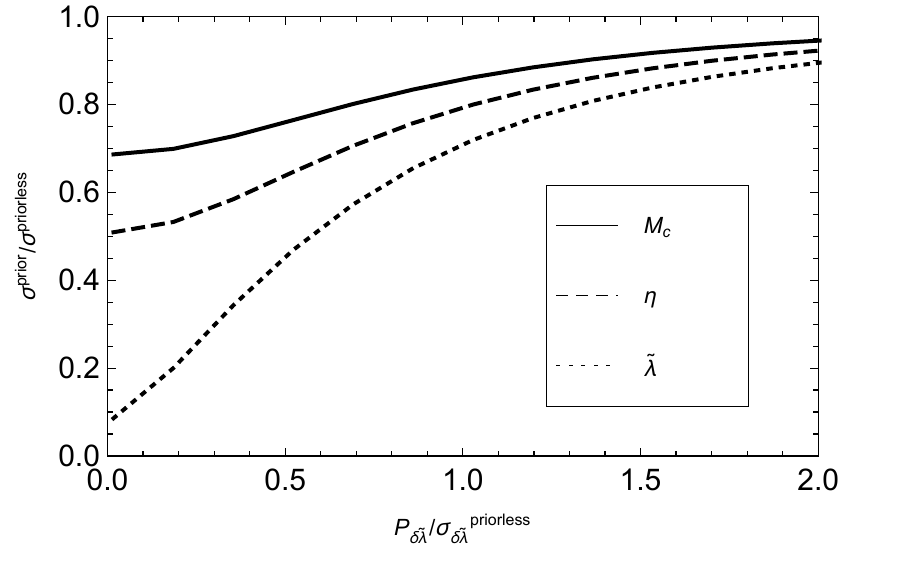}
         \caption{\label{fig.prior-error}   Contribution of the prior on $\delta \tilde{\lambda}$ to the measurement of the other intrinsic parameters. 
         Here, $\sigma^{\rm prior}$ indicates the error calculated with the prior information, while $\sigma^{\rm priorless}$ indicates the error calculated with no prior.
         Note that, this result is independent of the SNR \cite{Cho_2022}.}
\end{figure}

Figure \ref{fig.prior-error} describes how the prior information for the parameter $\delta \tilde{\lambda}$ affects the measurement of the other intrinsic parameters, 
showing the ratio between $\sigma^{\rm prior}$ and $\sigma^{\rm priorless}$
as a function of the fractional prior size $P_{\delta \tilde{\lambda}}/\sigma^{\rm priorless}$.
Here, $\sigma^{\rm prior}$ indicates the error calculated with the prior on $\delta \tilde{\lambda}$ 
with the variance $P_{\delta \tilde{\lambda}}$, while $\sigma^{\rm priorless}$ indicates the error calculated with no prior information.
The result shows that the contribution of the prior is the highest (lowest) for $\tilde{\lambda} \ (M_c)$.
For example, if the prior size is similar to the error size i.e., $P_{\delta \tilde{\lambda}} \sim \sigma^{\rm priorless}$, 
$\sigma^{\rm prior}$ can be $\sim 70 \ (85) \%$ of $\sigma^{\rm priorless}$ for $\tilde{\lambda} \ (M_c)$.
Note that, since both the error ($\sigma^{\rm prior}$) and the prior ($P_{\delta \tilde{\lambda}}$) 
are divided by $\sigma^{\rm priorless}$, this result is independent of the SNR (for more details, refer to Fig. 7 of \cite{Cho_2022}).

\begin{figure}[t]
    \centering
         \includegraphics[width=\columnwidth]{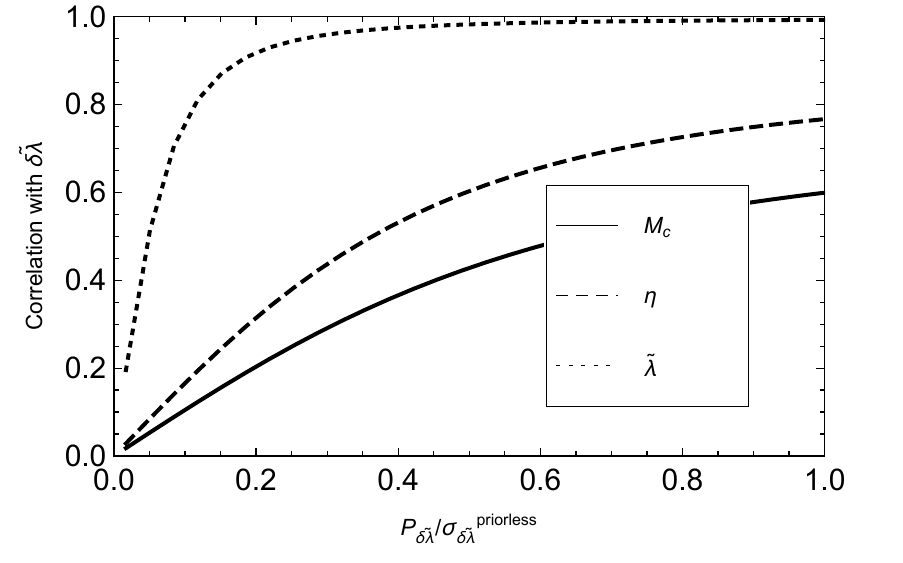}
     \caption{\label{fig.prior-correlation}  Dependence of the correlation on the prior. The curves represent the correlation coefficients between $\delta \tilde{\lambda}$ and the other parameters.}
\end{figure}

On the other hand, we also find in Table \ref{tab.FM-error} that the correlations between $\delta \tilde{\lambda}$ and the other parameters can be
reduced by incorporating the prior information.
In particular, the correlations with the mass parameters can be reduced to almost zero,
meaning that the parameter $\delta \tilde{\lambda}$ does not affect the measurement of the mass parameters.
To confirm this, we compute the 5--d Fisher matrix, in which $\delta \tilde{\lambda}$ is not included.
The result is given in the bottom row in Table \ref{tab.FM-error}.
We find that the errors of the mass parameters are nearly the same between the 5--d and the 6--d ($P_{\delta \tilde{\lambda}}=500$) Fisher matrices.
The dependence of the correlation on the prior is also described in Fig. \ref{fig.prior-correlation}, showing
the correlation coefficient as a function of $P_{\delta \tilde{\lambda}}/\sigma^{\rm priorless}$.

\subsection{Systematic bias due to eccentricity} \label{sec.bias}

\begin{figure}[t]
\begin{center}
\includegraphics[width=\columnwidth]{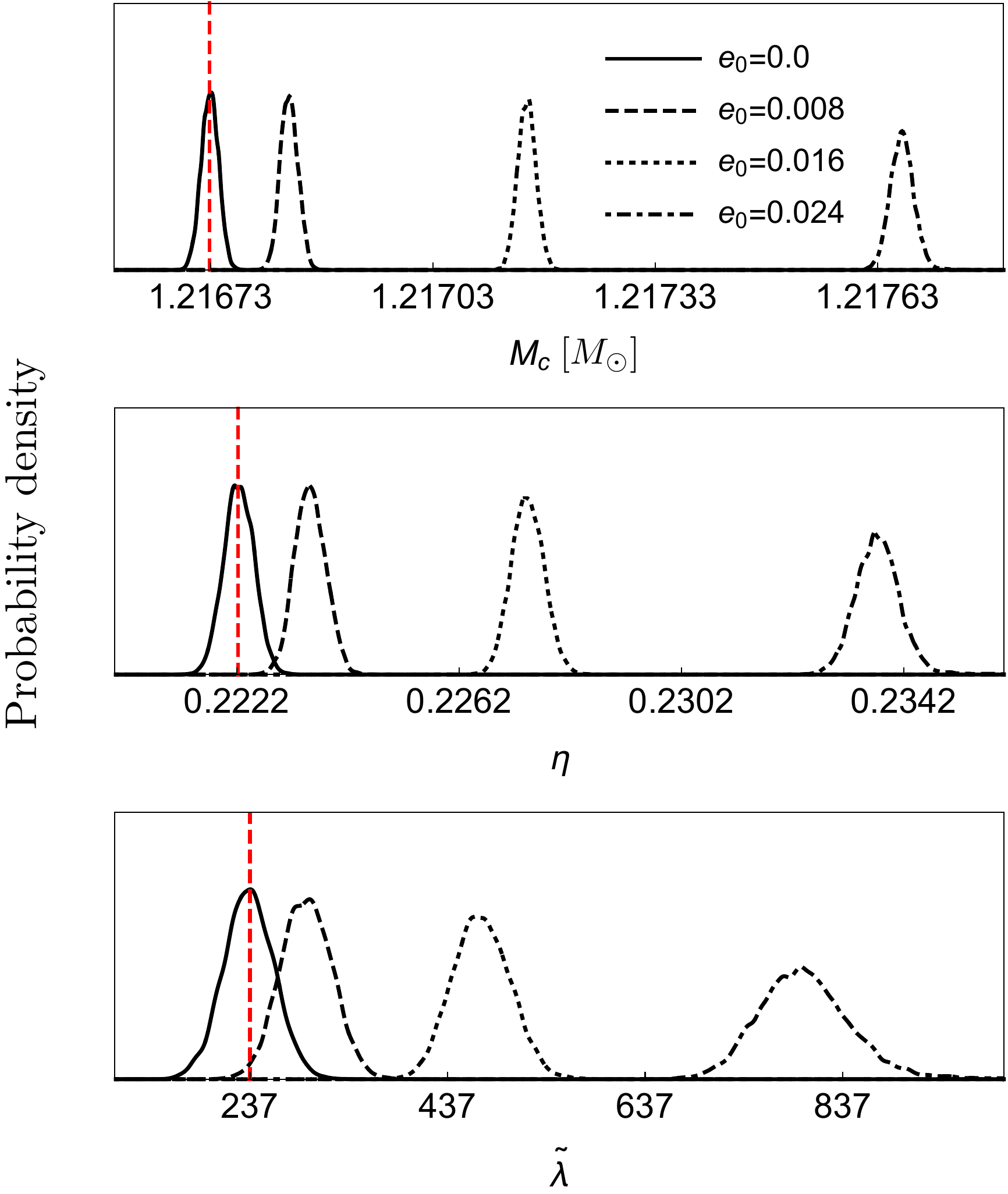}
\caption{\label{fig.1d-Mc-eta-lambda-biases} Marginalized PDFs for $M_c$ (top), $\eta$ (middle),
and $\tilde{\lambda}$ (bottom). We assume a high SNR ($\rho=114$) to remove unknow bias sources. 
For reference, the unbiased PDF is also given. The red dashed line represents the true value.}
\end{center}
\end{figure}

In this subsection, we investigate how the recovered parameters of eccentric BNS signals can be biased from their true values
by using circular waveforms in likelihood calculations.
To this end, we inject three BNS signals with small eccentricities $e_0=(0.008, 0.016,0.024)$
into the aLIGO PSD with zero noise.
Motivated by the results in Fig. \ref{fig.bilby-errors}, we assume a high SNR ($\rho=114$) to remove unknown bias sources.
Figure \ref{fig.1d-Mc-eta-lambda-biases} shows the marginalized posterior PDFs for the parameters $M_c$ (top), $\eta$ (middle),
and $\tilde{\lambda}$ (bottom), respectively.
As expected, as $e_0$ increases, the magnitude of the bias also increases. 
On the other hand, all of the biases show a positive direction, 
meaning that the recovered parameters are larger than the true parameter values.
This increasing behavior for the mass parameters is also seen in the BBH systems \cite{OShea:2021ugg,PhysRevD.105.023003}. 
In particular, the bias on the tidal parameter is important because the value of $\tilde{\lambda}$ can constrain the NS EOS.
For example, the soft EOS model ARP4 \cite{PhysRevD.79.124032}, 
which we employ in our analysis, gives $\tilde{\lambda} \sim 240$ for the binary with $(2\msun, 1\msun)$
but the stiffer EOS model MPA1 \cite{PhysRevD.79.124032} gives $\tilde{\lambda} \sim 450$.
Roughly speaking, if the astrophysical NSs are represented by the soft EOS,  
parameter estimation using circular waveform models for
eccentric BNS signals with $e_0 \gtrsim 0.016$ can yield results in favor of the stiff EOS models.
We will show a concrete example in Sec. \ref{sec.injection}.

\begin{table}[t]
\centering
\begin{tabular}{c|ccc|ccc}
\hline\hline
&\multicolumn{3}{c|}{Bayesian} & \multicolumn{3}{c}{FCV}\\
$e_0$ &  $ \Delta{M_c}/\sigma_{M_c}$ & $\Delta{\eta}/\sigma_{\eta}$ & $\Delta{\tilde{\lambda}}/\sigma_{\tilde{\lambda}}$ &  $ \Delta{M_c}/\sigma_{M_c}$ & $\Delta{\eta}/\sigma_{\eta}$ & $\Delta{\tilde{\lambda}}/\sigma_{\tilde{\lambda}}$ \\
\hline
0.008&2.841& 1.443 &0.685 &2.955&1.542 &0.696\\
0.016&11.42& 5.902 &2.634 &11.82& 6.170&2.784 \\
0.024&24.71& 12.84 & 6.531&26.60& 13.88&6.264\\
 \hline\hline
\end{tabular}
\caption{Comparison of biases between the Bayesian and the FCV methods. We use the errors ($\sigma$) given in Table \ref{tab.FM-error} with $P_{\delta \tilde{\lambda}}=500$.}
\label{tab.bilbyvsFM-bias}
\end{table}

\begin{figure}[t]
\begin{center}
\includegraphics[width=\columnwidth]{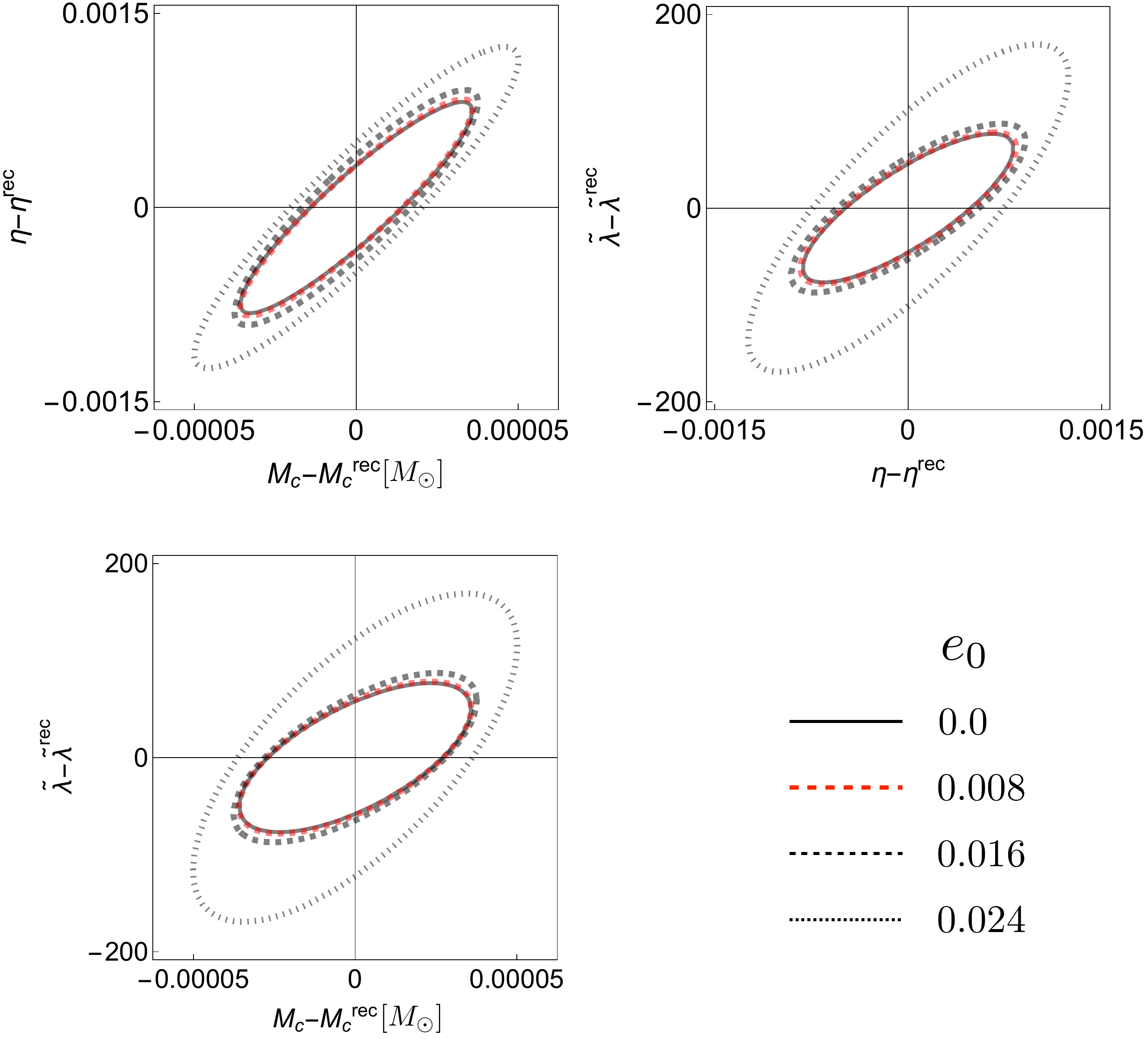}
\caption{\label{fig.2d-contours} $99 \% $ confidence regions of the biased posteriors. 
The recovered values ($\theta^{\rm rec}$) for the different signals coincide with the coordinates (0, 0). We assume $\rho=114$.}
\end{center}
\end{figure}

In Table \ref{tab.bilbyvsFM-bias}, we list the biases obtained 
from the Bayesian and the FCV methods for the eccentric signals given in Fig. \ref{fig.1d-Mc-eta-lambda-biases}.
For the FCV results, we use Eq. (\ref{eq.bias}) assuming the prior $P_{\delta \tilde{\lambda}}=500$. 
For efficiency, we give the fractional bias defined by $\Delta \theta / \sigma_{\theta}$ 
where $\sigma$ is the unbiased error given in Table \ref{tab.FM-error} with $P_{\delta \tilde{\lambda}}=500$.
Overall, the results of the FCV method are similar to those of the Bayesian method.
When $e_0=0.024$, the differences between the two methods are $\sim 7.6, 8.1,$ and $4.1 \%$ for $M_c, \eta,$ and $\tilde{\lambda}$, respectively.

On the other hand, one can see in Fig. \ref{fig.1d-Mc-eta-lambda-biases} 
that the PDF for $e_0=0.024$ is somewhat wider than the unbiased PDF, especially for $\tilde{\lambda}$.
This implies that a biased posterior can give a larger measurement error than the unbiased error.  
This is clearly illustrated in Fig. \ref{fig.2d-contours}, which shows the confidence regions of the posteriors together, 
where the recovered values ($\theta^{\rm rec}$) for the different signals coincide with the coordinates (0, 0).\footnote{In this plot, 
we used a covariance fitting function (implemented in {\bf Mathematica}) to the marginalized 2--d posteriors to obtain clear contours.}

\begin{figure}[t]
\begin{center}
\includegraphics[width=\columnwidth]{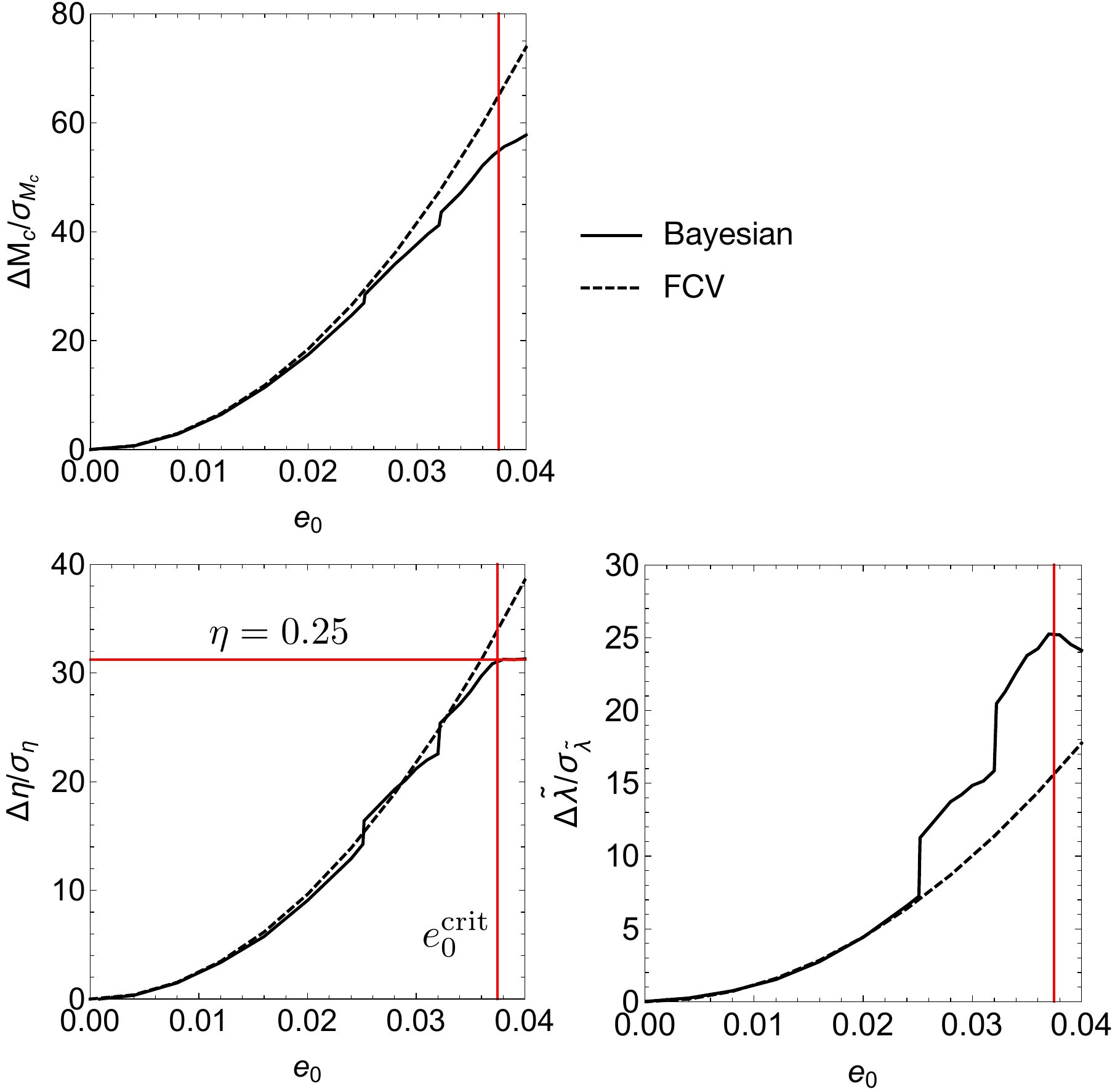}
\caption{\label{fig.bias-comparison-20Hz} Systematic bias due to eccentricity given as a function of $e_0$ obtained by using the Bayesian and the FCV methods. 
We present the fractional bias $\Delta \theta / \sigma_{\theta}$ where the measurement error is given Table \ref{tab.FM-error} with $P_{\delta \tilde{\lambda}}=500$.
The horizontal red line represents the upper boundary of $\eta$,
and the vertical red line indicates the value of $e_0$ when $\eta^{\rm rec}$ first touches the boundary.}
\end{center}
\end{figure}

\begin{figure*}[t]
\begin{center}
\includegraphics[width=2\columnwidth]{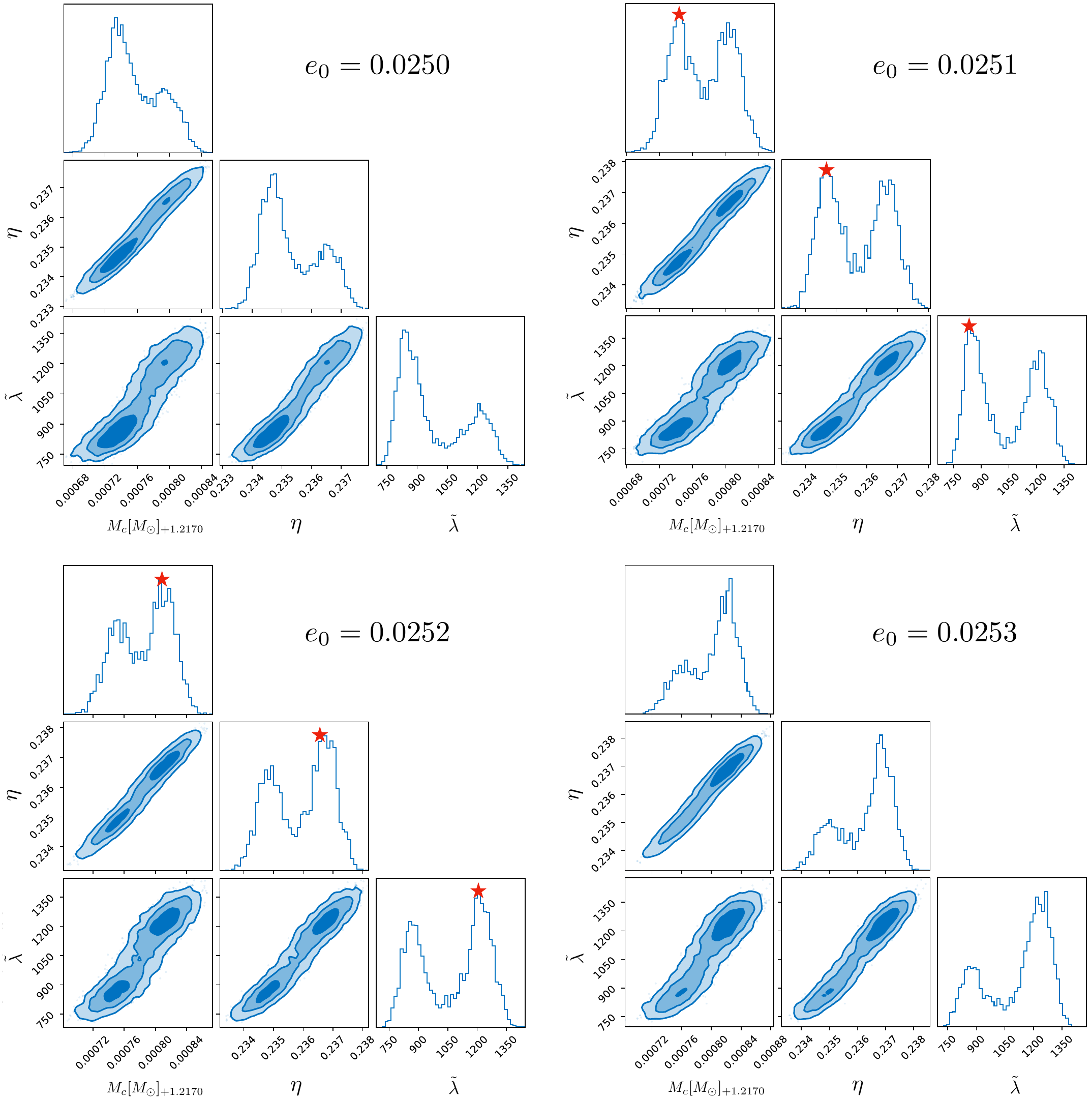}
\caption{\label{fig.bimodal} 
Biased posteriors showing the bimodality. We assume $\rho=114$. Note that, 
the maximum posterior position (marked with a red star) suddenly moves from the left peak to the right peak when $e_0$ is between 0.0251 and 0.0252.}
\end{center}
\end{figure*}

One can also see in Fig. \ref{fig.1d-Mc-eta-lambda-biases} that the increase in bias is faster for larger $e_0$.
To see a generalized tendency, we inject eccentric signals, varying $e_0$ up to 0.04, and perform parameter estimation.
In Fig. \ref{fig.bias-comparison-20Hz}, we display the bias as a function of $e_0$ for $M_c, \eta,$ and $\tilde{\lambda}$,
where we present together the bias calculated by the FCV method.
Favata \etal \cite{PhysRevD.105.023003} showed that the bias due to ignoring eccentricity
can be given as $\Delta \theta \propto e_0^2$.
We also find that our analytic curves can almost exactly match their quadratic fitting functions.

On the other hand, the Bayesian curves look similar to the analytic FCV curves up to $e_0 \sim 0.025$, but beyond that, they exhibit quite irregular behavior.  
One can see that the Bayesian curve rises abruptly at $e_0 \sim 0.025$ and $\sim 0.032$, which is especially noticeable for $\tilde{\lambda}$. 
To figure out how these sudden jumps occur, we generate a densely distributed signal set near $e_0=0.025$ and perform parameter estimation.
We choose four posterior results required to describe the sudden jump. 
We display the results for $e_0=(0.0250, 0.0251, 0.0252, 0.0252)$ in Fig. \ref{fig.bimodal}.
Interestingly, these posteriors show a bimodal distribution, and
we found that such bimodality only appears very near $e_0 \sim 0.025$ and $\sim 0.032$.
The maximum posterior position (marked with a red star) suddenly moves from the left peak to the right peak when $e_0$ is between 0.0251 and 0.0252,
resulting in a sudden jump in the bias curve.\footnote{Cho \cite{Cho_2015} also found sudden jumps in the bias curves for BBH systems (see Fig. 16 therein), 
in which the bias was produced by using an inspiral-only waveform model for the IMR signal.}
We also found similar situation near $e_0 \sim 0.032$.
The posterior's fine-scale structure becomes clearer if the SNR increases.
Thus, the bimodality will be more likely seen in the third-generation detectors, such as Einstein Telescope \cite{Punturo_2010} and Cosmic Explorer \cite{Abbott_2017}.

\begin{figure}[t]
\begin{center}
\includegraphics[width=\columnwidth]{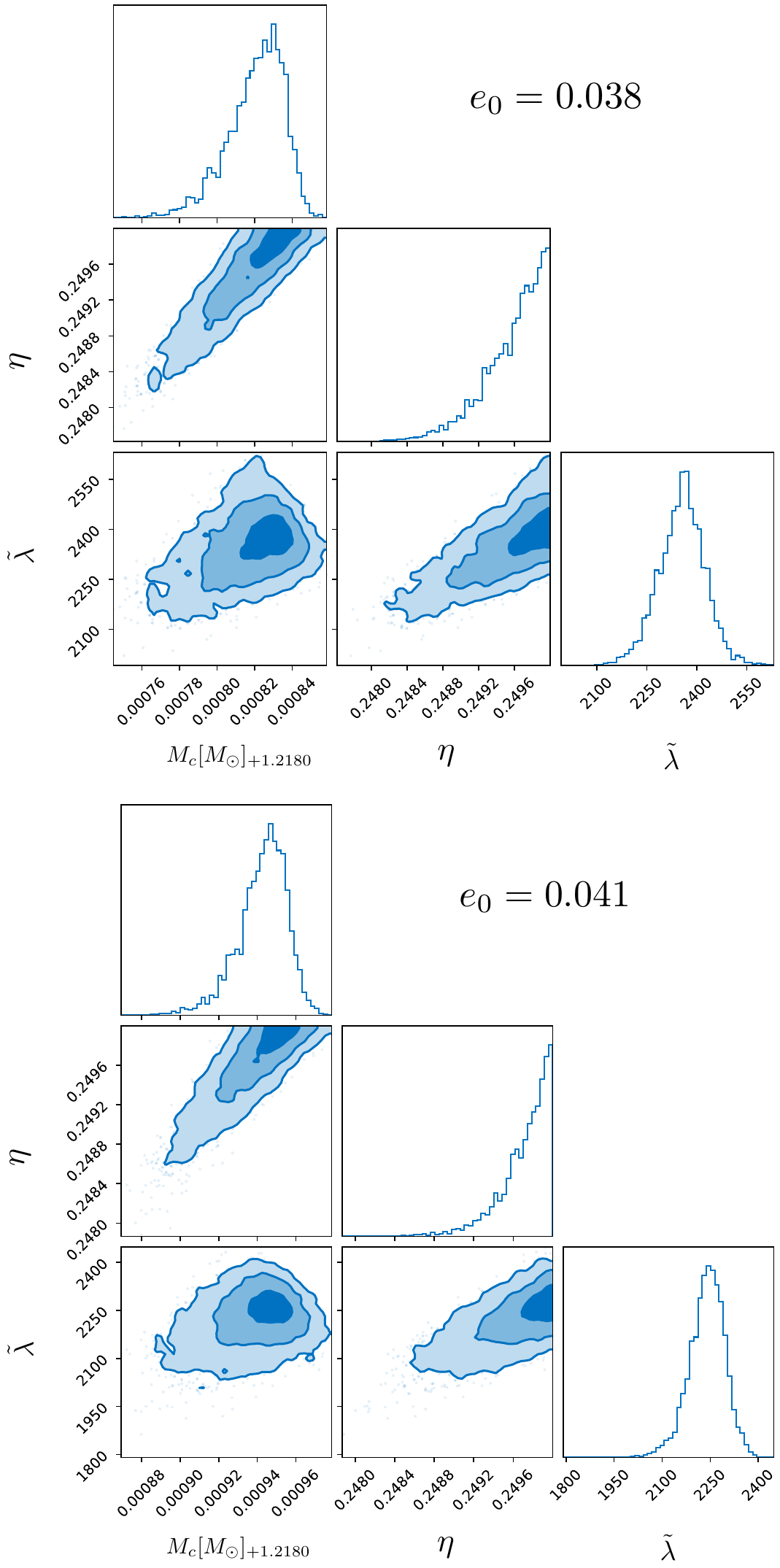}
\caption{\label{fig.eta-cut}Biased posteriors for the signals with $e_0=0.038$ (upper panel) and $e_0=0.041$ (lower panel), 
showing the increasing (decreasing) behavior for $\Delta M_c \ (\Delta \tilde{\lambda})$ in the region of $e_0 > e_0^{\rm crit}$. We assume $\rho=114$.}
\end{center}
\end{figure}

Although not the case for the chirp mass, the upper bound of the symmetric mass ratio is limited to $\eta \leq 0.25$.
Thus, the maximum bias for $\eta$ is given by $\Delta \eta=0.25-\eta_0$, independently of increasing $e_0$.
This relation implies that if the true value of $\eta$ is close to 0.25, 
the bias $\Delta \eta$ can be negligible compared to the measurement error.
By exploring the $M_c$--$\eta$ overlap surface for an equal-mass BNS source, 
Sun \etal \cite{PhysRevD.92.044034} also found ignorable biases even at large eccentricities of $\sim 0.4$.
In Fig. \ref{fig.bias-comparison-20Hz}, the horizontal red line represents the boundary $\eta=0.25$.
The vertical red line indicates the value of $e_0$ when $\eta^{\rm rec}$ first touches the boundary,
and we denote this critical eccentricity as $e_0^{\rm crit}$.
On the other hand, 
one can see that the increasing behavior of the bias curves changes significantly if $e_0$ increases beyond $e_0^{\rm crit}$.
The increase in $\Delta M_c$ slows down, $\Delta \eta$ remains at $0.25$,
and $\Delta \eta$ even decreases.
In Fig. \ref{fig.eta-cut}, we present two posterior results that illustrate such behavior well.
The result for $e_0=0.038$ corresponds to the case of $e_0 \sim e_0^{\rm crit}$,
and $e_0=0.041$ corresponds to $e_0 > e_0^{\rm crit}$.
In both cases, the 1--d PDFs for $\eta$ are truncated at the boundary and their peak positions coincide with the boundary, hence $\eta^{\rm rec}=0.25$.
The 2--d posteriors show that between $e_0=0.038$ and $e_0=0.041$, $\Delta M_c$ increases, but $ \Delta\tilde{\lambda}$ decreases. 
A detailed description is given in Appendix \ref{ap.eta-cut}

\subsection{Monte Carlo study}

\begin{figure}[t]
\begin{center}
\includegraphics[width=\columnwidth]{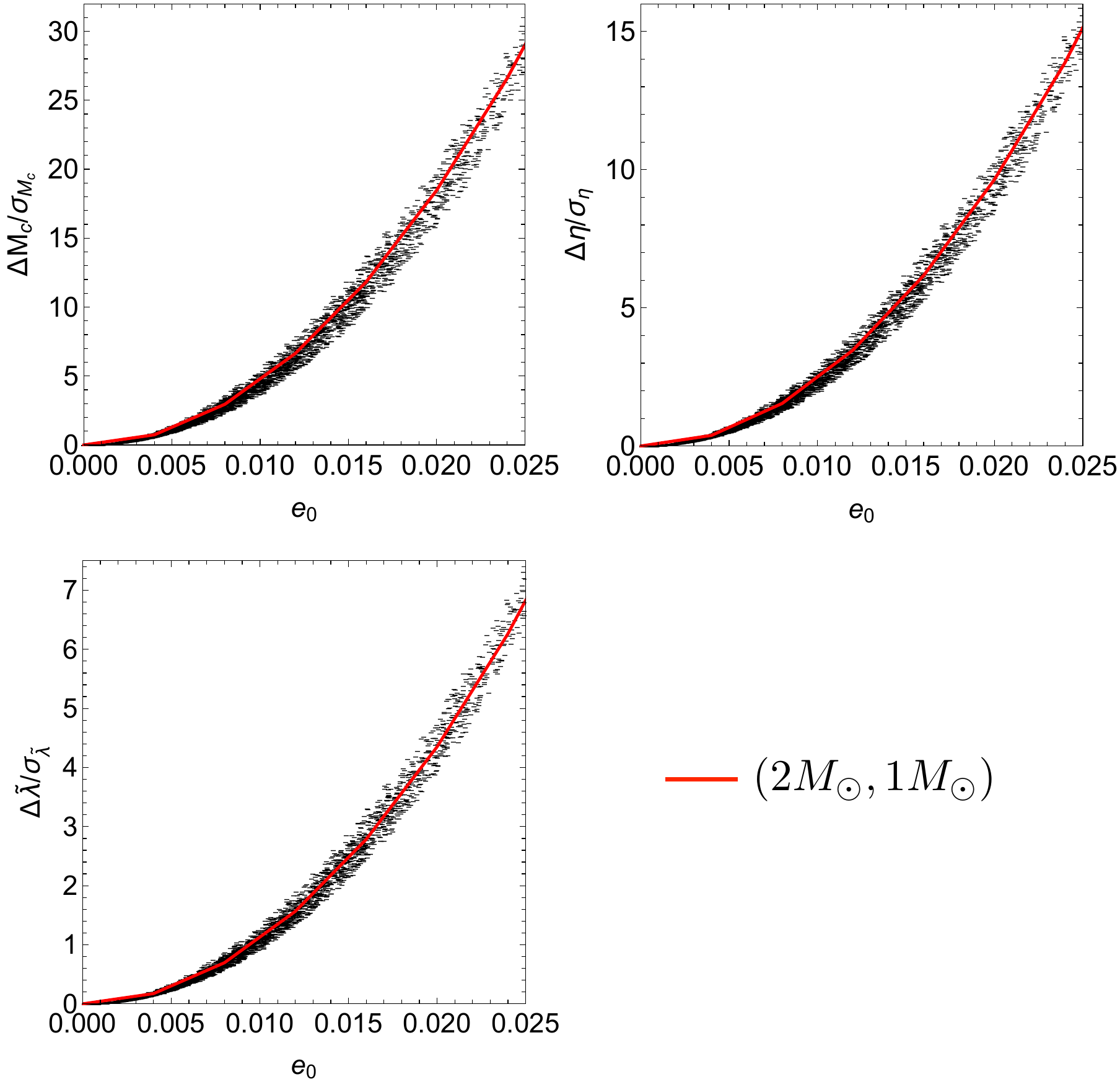}
\caption{\label{fig.bias-samples} Distribution of the fractional biases for the selected 4582 signals that satisfy $\eta^{\rm rec} \leq 0.25$. 
When calculating the analytic bias ($\Delta \theta$) and error ($\sigma_{\theta}$), 
we incorporate the prior $P_{\delta \tilde{\lambda}}=500$. The red line is the 1--d bias curve for the signal with masses of ($2\msun, 1\msun$).}
\end{center}
\end{figure}

We create $10^4$ Monte Carlo samples randomly generated in the parameter space $m_1$--$m_2$--$e_0$.
The parameter ranges are given as $1\msun \leq m_{1,2} \leq 2\msun \ (m_2 \leq m_1)$ and $0\leq e_0 \leq 0.025$.
Employing these samples as the true values, we produce $10^4$ eccentric signals and 
calculate biases using the FCV method assuming $P_{\delta \tilde{\lambda}}=500$.
We also calculate the unbiased errors using the same signals but assuming $e_0=0$ for all signals.
Finally, we select a total of 4582 signals that satisfy $\eta^{\rm rec} \leq 0.25$.
Figure \ref{fig.bias-samples} shows the distributions of the fractional biases for the selected signals.
For reference, the 1--d bias curves for the signal with masses of ($2\msun, 1\msun$) are given by the red lines.
For a given value of $e_0$, 
the width of the distribution represents the dependence on the component masses.
We find that the distribution shows a narrow band along the red line,
which means that the bias is mainly dependent on the value of $e_0$ and weakly dependent on the component masses.

\begin{figure}[t]
\begin{center}
\includegraphics[width=\columnwidth]{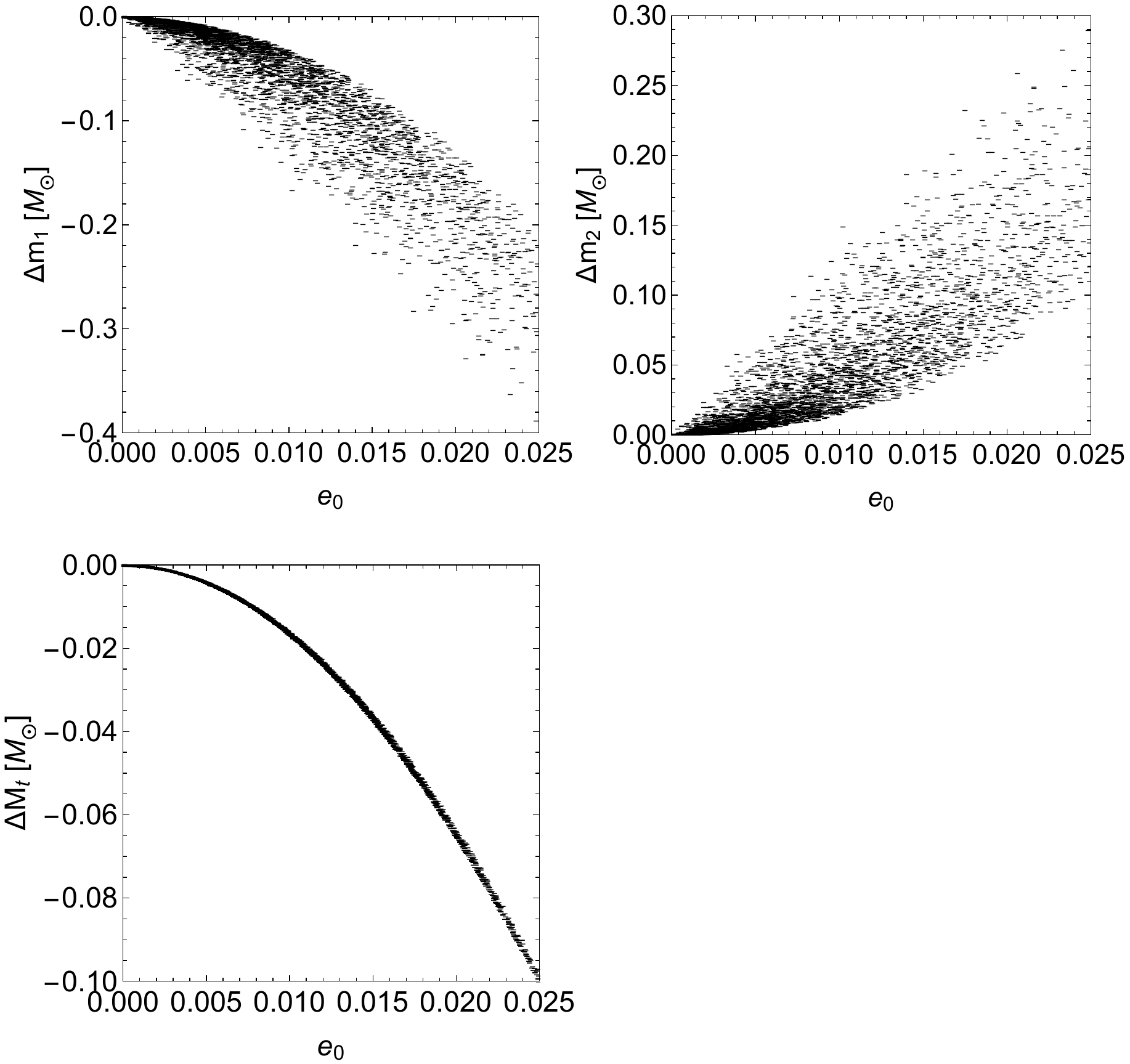}
\caption{\label{fig.bias-samples-2} Distribution of the biases for the same signals as in Fig. \ref{fig.bias-samples} 
but given for the component masses ($m_1, m_2$) and the total mass ($M_t$). 
The distribution for $M_t$ looks like a single curve, which means that the  distribution is almost independent of the component masses.
Note that, in this plot, we present only the bias, not the fractional bias.}
\end{center}
\end{figure}

To see a more useful explanation of the bias distribution,
we also display the biases for the component masses ($m_1, m_2$) and the total mass ($M_t$) in Fig. \ref{fig.bias-samples-2}.
Note that, in this plot, we present only the bias $\Delta \theta$, not the fractional bias $\Delta \theta / \sigma_{\theta}$.
The massive components have a negative bias, but the lighter components have a positive bias,
and both exhibit a wide distribution for a given value of $e_0$.
The maximum bias is given as $\Delta m_1 \ (\Delta m_2) \sim -0.4 \ (0.3) \msun$ at $e_0=0.025$.
For $M_t$, the biases are negative.
In particular, the bias distribution is almost independent of the signal mass,
so it even looks like a single curve.
The distribution for $M_t$ is quite consistent with the quadratic fitting function $\Delta M_t/\msun=-160 e_0^2$.

\subsection{Injection and recovery} \label{sec.injection}

\begin{figure}[t]
\begin{center}
\includegraphics[width=\columnwidth]{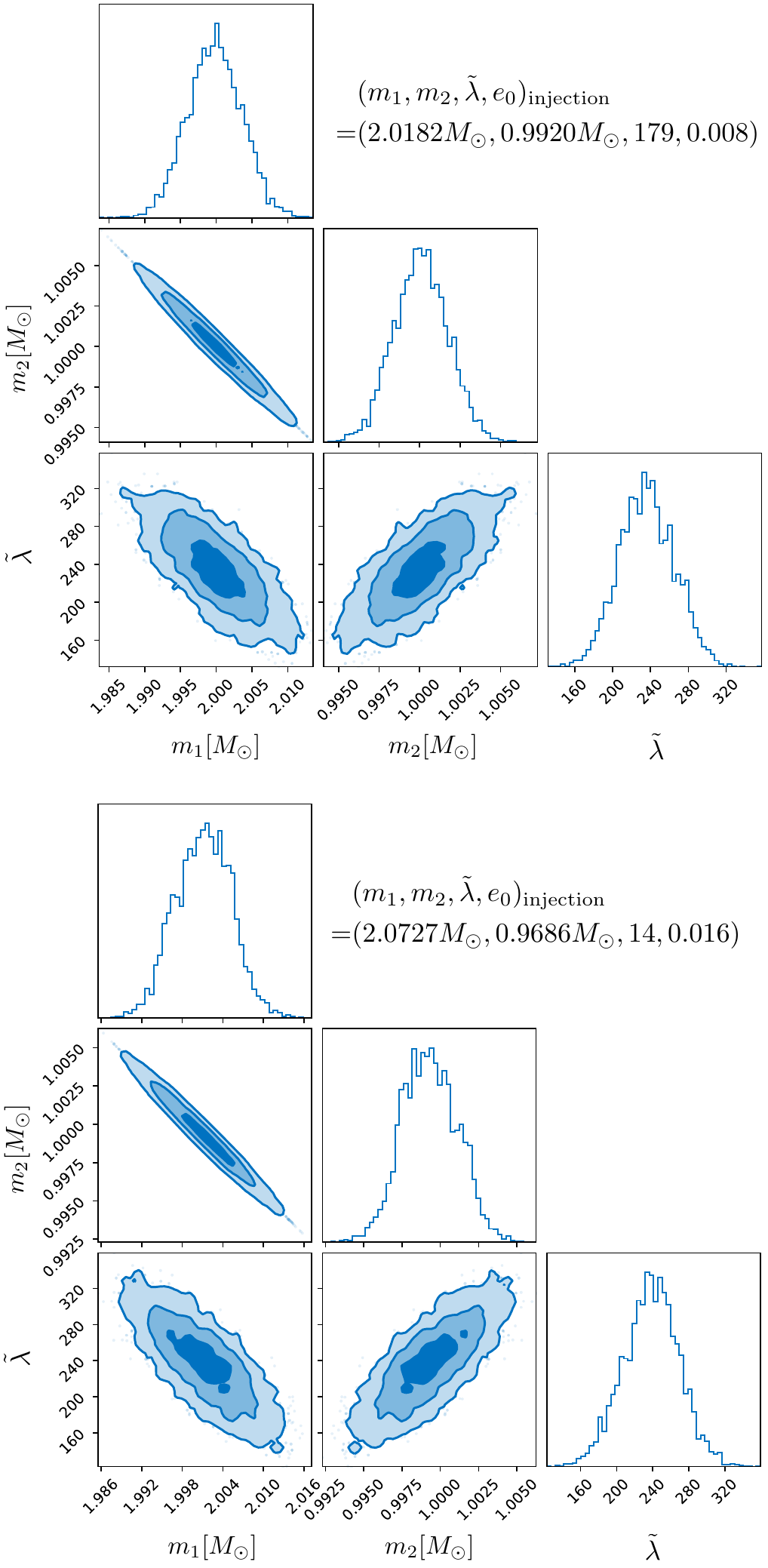}
\caption{\label{fig.recover} This example shows that different eccentric signals can be recovered as one identical binary source. 
The two posteriors have almost the same distribution and the same recovered parameters, i.e., $(m_1,m_2,\tilde{\lambda})\simeq(2\msun, 1\msun, 240)$.
We assume $\rho=114$.}
\end{center}
\end{figure}

In this subsection, we present two specific cases via injection and recovery.
First, we show how different eccentric signals can be recovered 
as one identical signal similar to our fiducial binary source i.e., $(m_1,m_2,\tilde{\lambda})\simeq(2\msun, 1\msun, 240)$.
To this end, we inject two signals with the true values of $(m_1,m_2,\tilde{\lambda},e_0)=(2.0182\msun, 0.9920\msun,179,0.008)$ 
and $(2.0727\msun, 0.9686\msun,14,0.016)$
and perform parameter estimation using circular waveforms.
The results are given in Fig. \ref{fig.recover}.
Although the signal in the lower panel has an $e_0$ value twice that of the signal in the upper panel,
their posteriors have almost the same distribution in the parameter space $m_1$--$m_2$--$\tilde{\lambda}$
and the same recovered parameters.
On the other hand, the biases of $M_t$ are $\Delta M_t \sim 0.010$ and $\sim 0.41$ for $e_0=0.008$ and 0.016, respectively.
Thus, the relation $\Delta M_t/\msun=-160 e_0^2$ is well satisfied.
Here, we estimated the injection parameters reversely from the fiducial values referring to the results in Fig. \ref{fig.bias-comparison-20Hz}.

\begin{figure}[t]
\begin{center}
\includegraphics[width=\columnwidth]{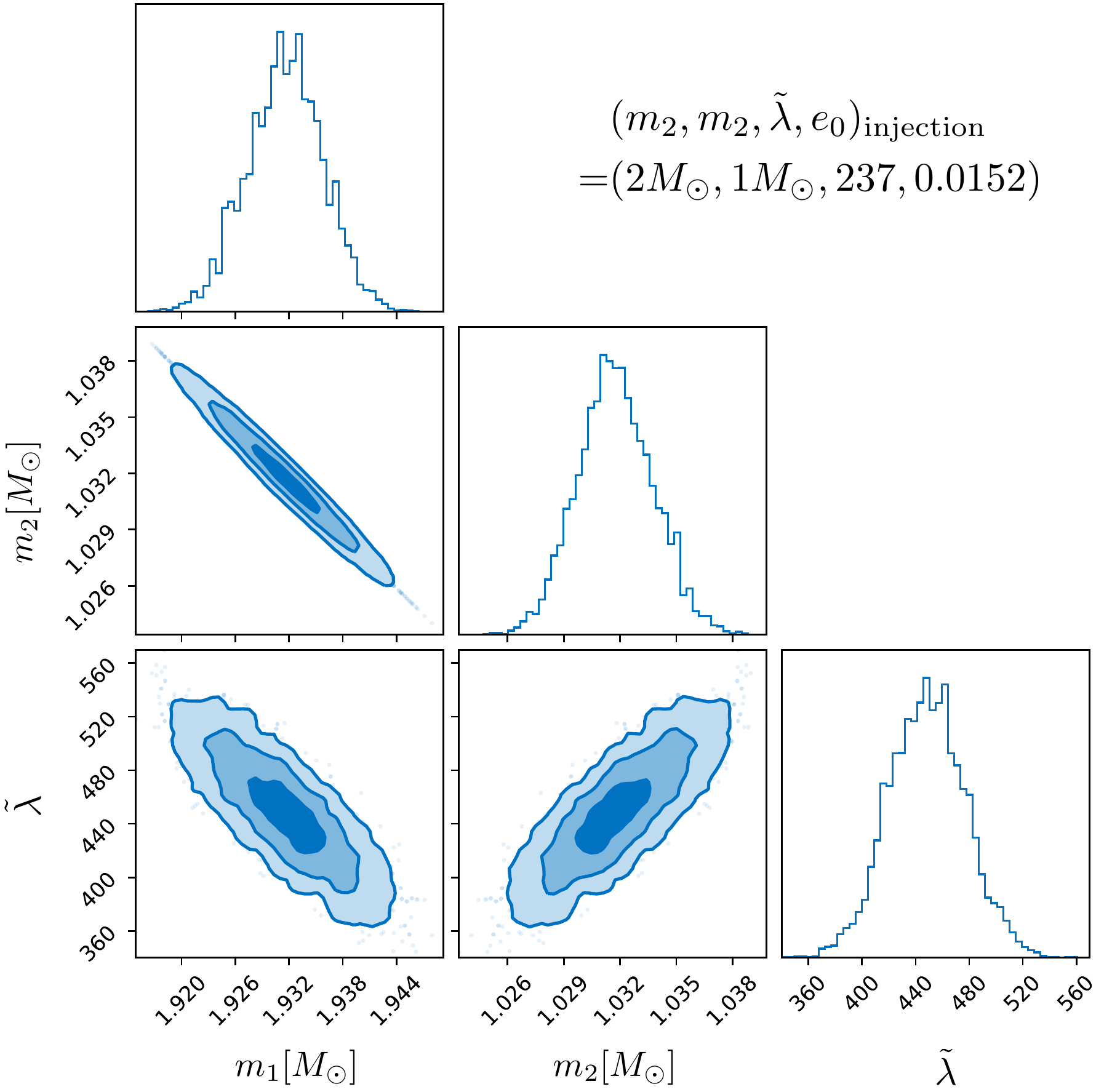}
\caption{\label{fig.e0-0.0152} This example shows that parameter estimation using circular waveforms can yield inaccurate NS EOS predictions.
The injection value ($\tilde{\lambda}=237$) is chosen according to the soft EOS model ARP4, 
but the recovered value ($\tilde{\lambda}^{\rm rec}\sim440$) representes the stiffer EOS (MPA1).
We assume $\rho=114$.}
\end{center}
\end{figure}

As discussed in Sec. \ref{sec.bias}, parameter estimation using circular waveforms for eccentric BNS signals
can yield incorrect predictions for the NS EOS.
Here, we present a concrete example.
We assume that the astrophysical NSs are represented by one of the soft EOS models APR4 \cite{PhysRevD.79.124032},
and then the tidal parameter value can be determined as $\tilde{\lambda} \sim 237$ for the given masses of $(2\msun, 1\msun)$. 
So, we inject a signal with the true values of $(m_1,m_2,\tilde{\lambda},e_0)=(2\msun, 1\msun,237,0.0152)$
and perform parameter estimation using circular waveforms.
The posterior result is given in Fig. \ref{fig.e0-0.0152}, where
the recovered parameters are $\sim (1.93\msun, 1.03\msun,440)$.
The tidal parameter of $\tilde{\lambda} \sim 440$
is consistent with the prediction of one of the stiff EOS models MPA1 \cite{PhysRevD.79.124032}. 
Therefore, this example demonstrates the importance of including eccentricity for the correct estimation of the NS EOS.

\section{Summary and discussion}\label{sec5}

In this work, we investigated how the small eccentricity of BNS signals
can affect parameter estimation for the mass parameters $M_c$ and $\eta$ and the tidal deformability $\tilde{\lambda}$.
We showed how the recovered parameters can be biased from their true values by presenting various posterior samples.
We also showed concrete posterior examples
that display irregular behavior of bias, such as the abrupt increase in bias.

Using the realistic population of BNS sources distributed in the $m_1$--$m_2$--$e_0$ space,
we calculated the measurement errors and the systematic biases for the parameters $M_c,\eta$, and $\tilde{\lambda}$
and obtained their generalized distributions in the range of $0 \leq e_0 \leq 0.025$.
For all of the three parameters, the fractional biases ($\Delta \theta/\sigma_{\theta}$) increase with increasing $e_0$,
and this increase is faster for larger $e_0$.
For a given value of $e_0$, the bias is weakly dependent on the masses and thus its distribution displays a relatively narrow band.
In particular, the bias distribution for the total mass exhibits a very narrow band, just like a single curve,
and can be fitted by the quadratic function $\Delta M_t/\msun=-160 e_0^2$.

We presented two specific injection-recovery examples.
The first example showed that independent parameter estimations for different signals with different eccentricities can give
almost identical recovered values for the component masses and $\tilde{\lambda}$.
In the second example, we showed that parameter estimation for signals with small eccentricity ($\sim 0.0152$) can yield predictions
that have shifted from the ``true" soft EOS to the ``false" stiff EOS, demonstrating the importance of including eccentricity in the waveform model.

We employed both the Bayesian and the analytic Fisher matrix methods for error calculations
and demonstrated that the analytic method can predict the parameter estimation errors well
by appropriately imposing the prior information.
Unlike the Fisher matrix, the FCV method has been used very limitedly in the literature
because its general validity has not been investigated.
We provided specific comparison results between the Bayesian and the FCV methods.
For the first time, we suggested the valid criteria of the FCV method by showing the bias as a function of the parameter that induces the bias.
In the era of the third-generation detectors,
the analytic methods will be more useful
because the length of the GW signal is greatly increased
and parameter estimation requires a very long computational time.
The impact of bias will also be more pronounced because 
the measurement errors can be reduced due to the increase in SNR by the third-generation detectors.
Meanwhile, the spin effect can be included in the Fisher matrix and the FCV method by employing the spinning TaylorF2 model,
and our approach will be extended to spinning binary systems in future works.


%

\begin{acknowledgments}
This work was supported by the National Research
Foundation of Korea (NRF) grants funded by the Korea
government (No. 2018R1D1A1B07048599, No. 2019R1I1A2A01041244, and No. 2020R1C1C1003250)
\end{acknowledgments}

\appendix

\section{PARAMETER ESTIMATION FOR THE ENTIRE 11 PARAMETERS}  \label{ap.full-param}

\begin{figure*}[t]
\begin{center}
\includegraphics[width=2\columnwidth]{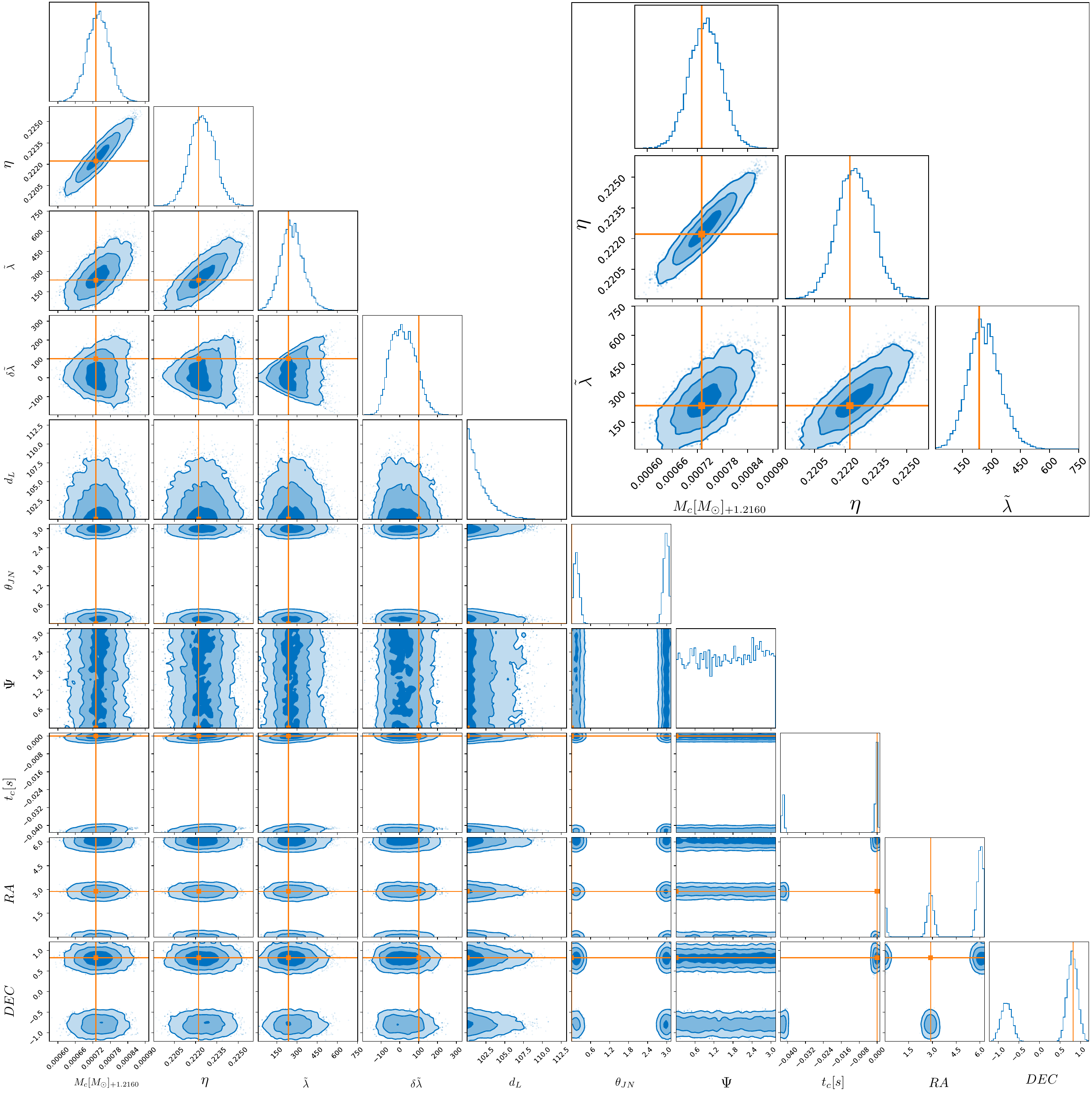}
\caption{\label{fig.full-param} Parameter estimation result for the same BNS source as in Fig. \ref{fig.bilby-errors} with $\rho=34$.
We use the entire 11 parameters including the 5 extrinsic parameters.
The zoom-in view of the posteriors for the three main parameters $M_c, \eta,$ and $\tilde{\lambda}$ is given in the inset. 
$\phi_c$ was automatically marginalized and thus is not shown here. True values are marked in orange.}
\end{center}
\end{figure*}

Here, we present the parameter estimation result for the entire 11 parameters 
using the same BNS source as in Fig. \ref{fig.bilby-errors}.
In the inset, we show the zoom-in view for the three main parameters $M_c, \eta,$ and $\tilde{\lambda}$.
We assume $\rho=34$, so this signal is the same as the one in Fig. \ref{fig.bilby-errors} with $\rho=34$.
It can be seen that the posteriors in the inset are well consistent with those in Fig. \ref{fig.bilby-errors}, including the small biases.
Thus, in a single-detector analysis, reducing the extrinsic parameters to a single parameter $\deff$
yields nearly the same parameter estimation results for the intrinsic parameters compared to using the entire parameters.
Since $\phi_c$ was analytically marginalized in the likelihood calculation process, it cannot be shown in the posterior.
The priors for the extrinsic parameters are given in default in {\bf Bilby}.

\section{CONFIDENCE REGION AT $e_0 \geq e_0^{\rm crit}$}  \label{ap.eta-cut}

\begin{figure}[t]
\begin{center}
\includegraphics[width=\columnwidth]{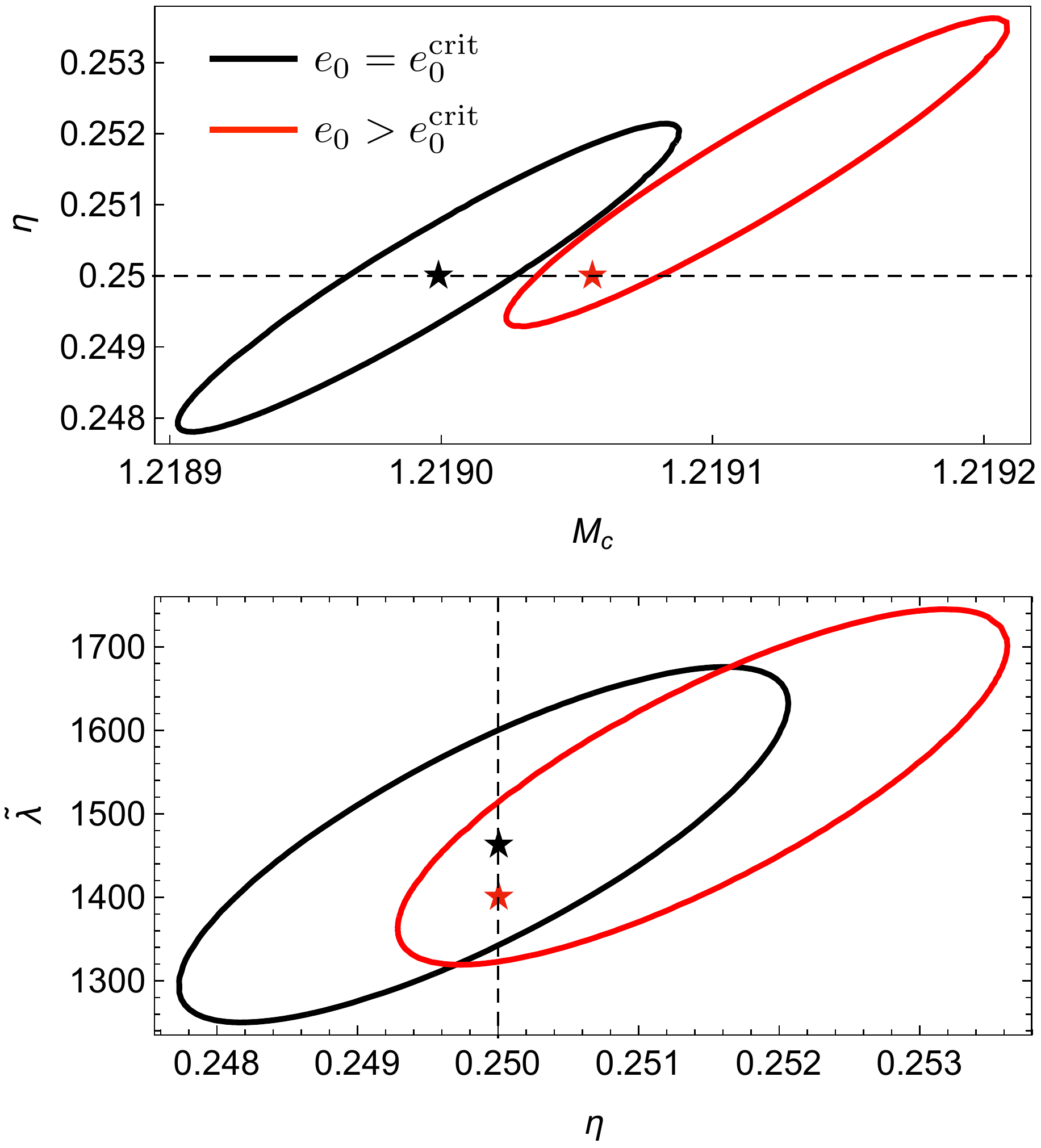}
\caption{\label{fig.eta-cut-bias} Overlap contours showing the increasing bias for $M_c$ (upper) and the decreasing bias for $\tilde{\lambda}$ (lower) in the region of $e_0 > e_0^{\rm crit}$.}
\end{center}
\end{figure}

Here, we show a schematic view describing the increasing bias for $M_c$ 
and the decreasing bias for $\tilde{\lambda}$ shown in Fig. \ref{fig.eta-cut} at $e_0 \geq e_0^{\rm crit}$.
Figure \ref{fig.eta-cut-bias} displays two overlap contours in the $M_c$--$\eta$ (upper) and $\eta$--$\tilde{\lambda}$ planes, respectively.
These biased contours can be obtained by using the Fisher matrix and the FCV methods for our fiducial BNS source.
The maximum overlap position ($P_{\rm max}$) is located at the center of the contour.
The star indicates the maximum overlap position only in the region below the boundary $\eta=0.25$.
In the upper panel, the difference between the red and the black stars is smaller than
the difference between $P_{\rm max}$ of the red contour and $P_{\rm max}$ of the black contour.
This means that the increase in $\Delta M_c$ slows down.
In the lower panel,
although $P_{\rm max}$ of the red contour is higher than $P_{\rm max}$ of the black contour,
the red star is lower than the black star,
and that indicates the decrease in $\Delta \tilde{\lambda}$.
Note that, these contours are comparable to the confidence regions in Fig. \ref{fig.eta-cut},
except that these contours are extended beyond the $\eta$--boundary.

\section{DEPENDENCE ON THE TRUE VALUES OF $\tilde{\lambda}$ AND $\delta \tilde{\lambda}$}  \label{ap.tidal-dependence}
\begin{figure}[t]
\begin{center}
\includegraphics[width=\columnwidth]{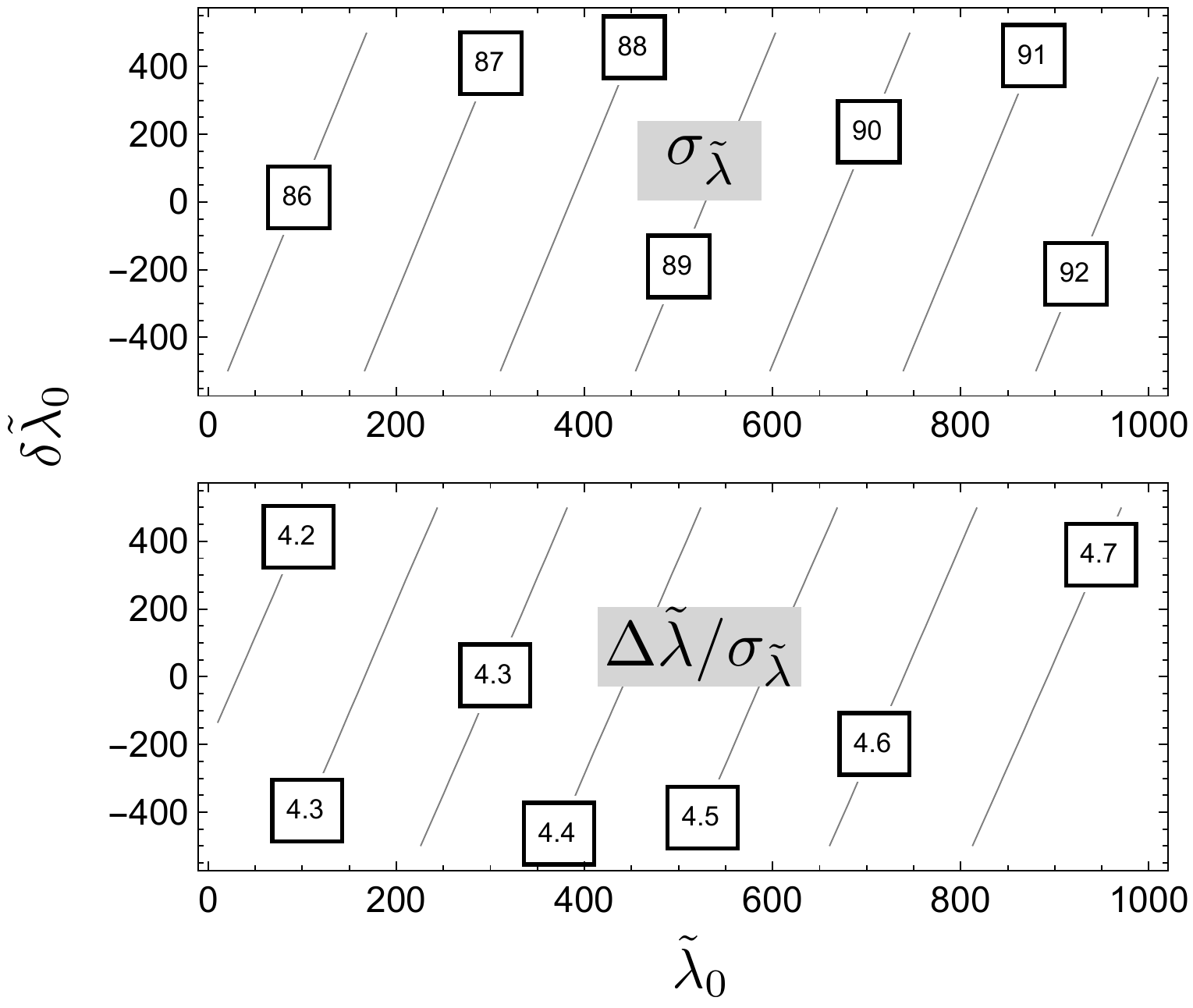}
\caption{\label{fig.lambda-true-dependence}Errors and fractional bias for $\tilde{\lambda}$ calculated by varying the true values of the tidal parameters.
The other true values are given as $(m_1,m_2,e_0)=(2\msun,1\msun,0.02)$.
We assume $\rho=34$.}
\end{center}
\end{figure}

Here, we show how
the true values of $\tilde{\lambda}$ and $\delta \tilde{\lambda}$ affect our results of errors and biases.
We assume a BNS signal with the true values of $(m_1,m_2,e_0)=(2\msun,1\msun,0.02)$.
Using the Fisher matrix and the FCV methods, 
we calculate the errors and the biases for the parameters $M_c,\eta,$ and $\tilde{\lambda}$ varying the true values of the tidal parameters $\tilde{\lambda}$ and $\delta \tilde{\lambda}$.
We find that the results for the mass parameters are independent of the tidal parameters, so they are not shown here.
The errors and the biases for $\tilde{\lambda}$ are given in the upper and the lower panels, respectively, in Fig. \ref{fig.lambda-true-dependence}.
It can be seen that the dependence on $\delta \tilde{\lambda}$ is almost negligible.
Although the dependence on $\tilde{\lambda}$ is relatively large, 
the variation in the results is less than $10\%$ in our parameter range.
For example, for the soft ARP4 EOS model, $\sigma_{\tilde{\lambda}} = 86.9, \Delta{\tilde{\lambda}}=544.9$,
and for the stiffer MPA1 EOS model, $\sigma_{\tilde{\lambda}} = 88.4, \Delta{\tilde{\lambda}}=569.1$.
Note that, for $M_c$ and $\eta$, the differences between the two EOS models are less than $0.01\%$.

%

\newpage

\bibliographystyle{apsrev4-2}

\bibliography{biblio}


\end{document}